\documentclass[%
 reprint,
superscriptaddress,
 amsmath,amssymb,
 aps,
]{revtex4-2}

\usepackage{graphicx}
\usepackage{dcolumn}
\usepackage{bm}
\usepackage{tikz}
\usetikzlibrary{arrows}
\usepackage{subfigure}
\usepackage{amsmath}
\usepackage{amssymb}
\usetikzlibrary{shapes,arrows, decorations.text}
\usetikzlibrary{external}
\usepackage{hyperref}
\usepackage{lineno}
\hypersetup{
    colorlinks=true
}

\begin{document}

\preprint{APS/123-QED}

\title{Validation of straight-line signal propagation for radio signals of very inclined cosmic ray air showers}

\author{Dieder Van den Broeck}
\email{Dieder.jan.van.den.broeck@vub.be}
\affiliation{Vrije Universiteit Brussel (VUB),
 Pleinlaan 2, Brussels, Belgium}

\author{Uzair Abdul Latif}
\affiliation{Vrije Universiteit Brussel (VUB),
 Pleinlaan 2, Brussels, Belgium}

\author{Stijn Buitink}
\affiliation{Vrije Universiteit Brussel (VUB),
 Pleinlaan 2, Brussels, Belgium}
 
\author{Krijn de Vries}
\affiliation{Vrije Universiteit Brussel (VUB),
 Pleinlaan 2, Brussels, Belgium}

\author{Tim Huege}
\affiliation{Karlsruhe Institute of Technology (KIT), PO Box 3640, 76021 Karlsruhe, Germany}
\affiliation{Vrije Universiteit Brussel (VUB), 
 Pleinlaan 2, Brussels, Belgium}

\date{\today}

\begin{abstract}
An ongoing challenge for radio-based detectors of high-energy cosmic particles is the accurate description of radio signal propagation in natural nonuniform media.
For radio signals originating from extensive air showers, the current state of the art simulations often implicitly assume straight-line signal propagation. The refraction due to a nonuniform atmosphere is however expected to have an effect on the received signal and associated reconstruction. This effect is currently not completely understood for the most inclined geometries.
Here, we present a study regarding the validity of straight-line signal propagation when simulating radio emission associated with very inclined air shower geometries. To this end, the calculation of the electric field based on the end point formalism used in CoREAS was improved by use of tabulated ray tracing data. We find a difference of $2\%$ in radiation energy and a difference of $O(0.01^{\circ})$ on direction reconstruction when working at frequencies below $600$ MHz. We thus find that, for frequencies up to $600$ MHz and zenith angles up to $88^{\circ}$, the current straight-line based simulation approaches are accurate.  
\end{abstract}

\maketitle

\section{Introduction}
To study and understand the highest energy processes in the Universe, astroparticle physics aims to detect the high energy particles that originate in these processes. To achieve this, many different detection efforts have been designed and constructed to detect different cosmic messengers. For cosmic rays, the current highest energy results are reported by the Pierre Auger observatory~\cite{PierreAuger:2021hun} and Telescope Array observatory~\cite{TelescopeArray:2012vhh} with energies up to around $10^{20}$ eV. For neutrinos, IceCube has reported a diffuse neutrino flux around $10^{15}$ eV~\cite{IceCube:2020wum}.
\\
\par
At high energies, cosmic particles can be detected by measuring the particle cascades that they produce upon interaction in media such as the Earth's atmosphere or polar ice.
Since fluxes drop rapidly at the highest energies, the need arises for large detection volumes. To this end, radio based detectors can play a crucial role. For in-ice geometries, radio antennas provide a relatively inexpensive method to potentially probe large effective volumes. The general low attenuation of radio signals in ice is one of the key reasons for the current interest in radio detection methods for high-energy cosmic particles in ice ~\cite{Barrella:2010vs,Barwick:2005zz,Besson:2007jja,Avva:2014ena}. Recent advancements in understanding the emission coming from particle cascades allows for radio detection to be used to probe in air particle cascades~\cite{Huege:2016veh}. 
\\
\par
An example of a detection effort that makes use of radio-based detection is the AugerPrime upgrade~\cite{AugerPrimeUpgrade}. This upgrade of the Pierre Auger observatory adds an array of radio antennas complementing the Auger detector strategy to detect extensive air showers induced by ultrahigh energy cosmic rays (UHECR)~\cite{Huege:2023pfb}. Important for the context of this work is that the radio part of Auger will focus on detecting the most inclined showers, due to the large associated radio footprint.
The Giant Radio Array for Neutrino Detection (GRAND) is a planned detector which aims to detect extensive air showers using radio antennas, with particular interest to showers induced by a $\tau$-particle originating from a $\tau$-neutrino interaction in surrounding mountains~\cite{GRAND:2018iaj}.
A similar detection strategy is envisioned for the Beamforming Elevated Array for Cosmic Neutrinos (BEACON)~\cite{Southall:2022yil}, which aims to detect high energy $\tau$ neutrinos interacting in the earth and inducing an upward-going particle cascade.
Other experiments to consider are TAROGE~\cite{TAROGE} and the associated TAROGE-M~\cite{TAROGE-M:2022soh} observatories. These experiments are situated atop elevated surfaces, such as mountains, and near seawater. The ocean surface is then used as a reflector for radio signals to allow for the detection of high-energy particle cascades. 
All of these observation efforts focus on very inclined geometries, where the shower propagates nearly horizontally. 
\\
\par
To calculate the radio emission coming from extensive air showers (EAS), one often relies on particle level simulation software. Some notable examples are CoREAS~\cite{Huege:2013vt}, based on the endpoint formalism, and ZHAireS~\cite{Alvarez-Muniz:2010hbb}, based on the ZHS formalism. These simulation codes are time optimized, and as such rely on assumptions which might no longer be valid when moving from vertical to more inclined geometries. One of these assumptions, and also the focus of this work, is the assumption of straight-line signal propagation. When working with a nonuniform atmosphere, the ray tracing paths that one would use to describe signal propagation are in general curved. This curvature of the ray paths is more pronounced for very inclined geometries, where showers develop at higher altitude, causing a larger difference in the index of refraction values between emitter and observer. As will be discussed, this assumption plays a role in the implementation of geometric boosting, which is used to include Cherenkov-like effects.
\\
\par
It is well known that the travel time difference between curved, fully raytraced, paths and straight line paths that use an average index of refraction is small, of the order of $O(0.1)$ ns \cite{Alvarez-Muniz:2015ayz,Schluter:2020tdz,Werner:2007kh}.
This already provides a qualitative argument that moving from a straight line to a curved ray path should not radically alter the region over which one would observe coherent radiation, due to the introduced extra phase being negligible.
In \cite{Alvarez-Muniz:2015ayz}, a model to describe radio pulses is also used to show that the effect on the observed amplitude remains minimal when compared to the standard ZHAireS implementation for frequencies up to $900$ MHz and zenith angles up to $85^{\circ}$. Aside from raytracing, other methods exist to describe signal propagation in nonuniform media. Some examples include finite difference time domain methods~\cite{Deaconu:2018bkf}, parabolic equation solvers~\cite{RadarEchoTelescope:2020nhe}, and recently an approach that reconstructs the radio signal by characterizing the ice by use of Green's functions~\cite{Windischhofer:2023ahw}. These methods give higher accuracy results, but are typically also more computationally intensive than ray tracing.
\\
\par
While the small time difference between curved and straight-line propagation is well understood, a full description of how to account for the effect of ray curvature in current simulation codes has not yet emerged. This is crucial, as future experiments plan to operate within media with large index of refraction gradients~\cite{RNO-G,GRAND:2018iaj,Southall:2022yil} where the difference in travel time between a straight and curved path is expected to become more substantial, especially for the modeling of radiation coming from air showers that develop partially in more dense media \cite{DeKockere:2024qmc}.
\\
\par
In this work, we present a study regarding the effects of the straight line approximation commonly used for the calculation of radio signals coming from air showers, focusing on the effects for very inclined air showers. We first present the study of differences in travel time due to ray curvature before moving on to the description of Cherenkov-like effects by the geometric boostfactor and the current parametrization of this boostfactor in CoREAS. This study aims to determine if there exist geometries for which the straight line approximation used in many state-of-the-art simulation codes would potentially cause a significant uncertainty.

\section{CoREAS and the end point formalism}
The end-point formalism calculates the emission from accelerated particles by dividing the trajectory of the emitter as a collection of straight line tracks. The particle is then instantly accelerated and decelerated at the so-called end points of these tracks \cite{EndPoints}. This end point formalism is at the base of the CoREAS code which is implemented in the CORSIKA software package to allow for the simulation of radio emission coming from extensive air showers. In CoREAS the induced electric field contributions are calculated as~\cite{EndPoints,Ludwig:2010pf}:
\begin{eqnarray}\label{eq:End-point}
\vec{E}_{\pm}(\vec{x},t) = \pm \frac{1}{\Delta t} \frac{q}{c}\left( \frac{\hat{r} \times [\hat{r} \times \vec{\beta}^*]}{|1-n\vec{\beta}^* \cdot \hat{r}|R} \right),
\end{eqnarray}
where $\vec{\beta}^*$ is the ratio between the velocity of the emitting charge and the speed of light in vacuum, $\hat{r}$ is the unit vector along the direction connecting emitter and receiver, $R$ is the geometrical distance between emitter and receiver and $\Delta t$ is the observer time window.
The denominator $|1-n\vec{\beta}^* \cdot \hat{r}|$ arises from a term $\frac{\mathrm{d}t}{\mathrm{d}t'}$\cite{EndPoints} and serves as the definition for the geometrical boostfactor $B_f$. This boostfactor has to be included as, due to the index of refraction, it is possible for a signal emitted in an emission time interval $\Delta t'$ to arrive at the receiver in a shorter arrival time interval $\Delta t$. This boosting of the signal, purely due to geometry, can be described by the geometric boostfactor: $\left|\frac{\mathrm{d}t}{\mathrm{d}t'}\right|$. The current implementation calculates this boost factor by usage of a straight-line approximation for $\hat{r}$. This could potentially introduce an uncertainty in the simulated electric field, as will be discussed in the next section. 
\\
\par
Note that the current implementation of the endpoint formalism in CoREAS does take into account the average index of refraction  when calculating the travel time along the straight line path between a given source and observer. This is an important piece of information, as it was shown in \cite{Schluter:2020tdz} that refractive effects are already included in the standard CoREAS implementation of the end point formalism, even with the straight line approximation used when calculating the boostfactor. It is expected that these effects arise in the calculation because CoREAS takes into account the average index of refraction value along the straight line propagation path. As will be further discussed in Sec.~\ref{sec:TravelTime}, the difference for the travel time by assuming straight-line propagation in this manner is of the order $0.1$ ns for zenith angles up to $85^{\circ}$, and therefore the region where signal adds coherently shifts as if due to refractive effects.

\section{Ray tracing and the geometric Boostfactor}
A numerical approach to ray tracing can be found starting from Fermat's principle, which defines the ray paths by minimizing the optical path length (OPL) along the path. The optical path length $L$ for the raypath $\boldsymbol{r}$ between points $A$ and $B$ can be defined as \cite{holm2011geometric}
\begin{align}
    L=\int_A^B n(\boldsymbol{r}(s)) \cdot \mathrm{d}s,
\end{align} 
where $\mathrm{d}s^2=\mathrm{d}\boldsymbol{r}(s)\cdot \mathrm{d}\boldsymbol{r}(s)$.
Minimizing the variation in optical path length and imposing a two dimensional solution in the $(x,z)$ plane leads to the following set of equations:
\begin{align*}
    \dot{x}=\frac{p_x}{n(x,z)} && \dot{z}=\frac{p_z}{n(x,z)}
    \\
    \dot{p}_x= \frac{\partial n}{\partial x} && \dot{p}_z=\frac{\partial n}{\partial z}
\end{align*}
where the optical momenta are defined as $p_i= n \dot{x}$, $n(x,z)$ represents the index of refraction evaluated at coordinates $(x,z)$ and the dot derivative denotes the derivative with respect to the step $s$ along the path. This solution can now be paired with any differentiable index of refraction profile for $n(x,z)$ along with a suitable range of initial conditions to perform ray-tracing. For the study contained in this work, the Malargue (January) atmosphere found in the CORSIKA documentation~\cite{CorsikaWebsite} was used. Also note that a curved atmosphere is utilized, where the index of refraction is a function of altitude above sea level.
\\
\par
\begin{figure}
    \centering
    \includegraphics[width=\linewidth]{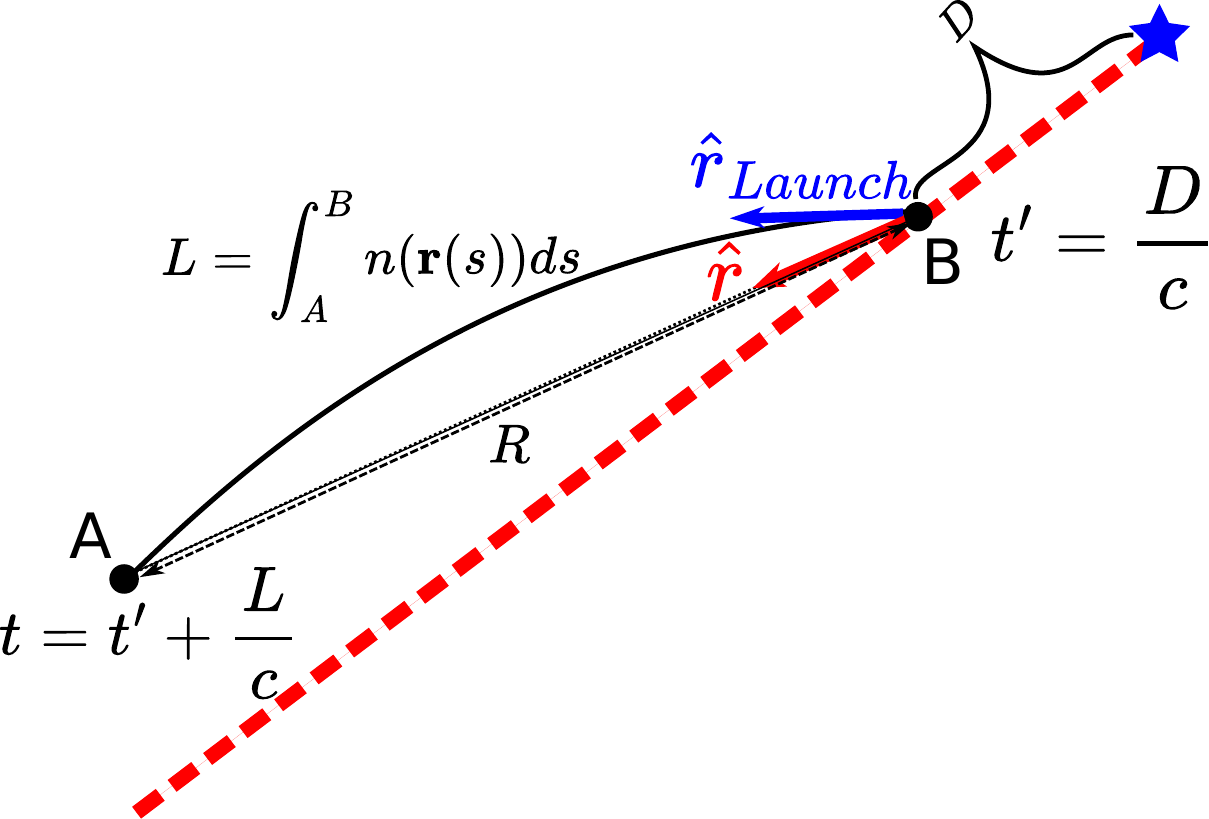}
    \caption{A diagram showing the line model of the cascade with important parameters highlighted. The first interaction point is marked by the blue star along with a receiver point A and an emission point B. The travel time is calculated from the optical distance of the ray. Also highlighted are the initial launch vector of the ray $\hat{r}_{\mathrm{Launch}}$ and vector $\hat{r}$ pointing from emitter to receiver along a straight-line. }
    \label{fig:RTParams}
\end{figure}

To investigate geometric boosting for emission coming from particle cascades, we modeled the cascade as emitters on a line.  For this line model, we define the emission time $t'$ and arrival time $t$ as
\begin{align*}
  t'=\frac{D}{c} ,
  \\
  t= t' + \frac{L}{c}
\end{align*}
With $D$ the distance to the cascade front measured from the first interaction point along the cascade propagation axis and c the speed of light in vacuum. A visualization is given in Fig.~\ref{fig:RTParams}, with important parameters highlighted.
For a uniform index of refraction, one can prove that the boostfactor can be written as:
\begin{align}
    B_f=\left|\frac{\mathrm{d}t}{\mathrm{d}t'}\right|=1 - n \vec{\beta}^* \cdot \hat{r}
    \label{eq:boostfactor_uni}
\end{align}
with $\Vec{\beta}^*$ and $\hat{r}$ defined as in the end point formalism. The link to boosting becomes clear when one considers the extreme case of maximum boosting where signal emitted over a finite emission time interval $\mathrm{d}t'$ arrives in an infinitesimal arrival time interval $\mathrm{d}t$ as then $\frac{\mathrm{d}t}{\mathrm{d}t'} = 0$. In this, case Eq.(~\ref{eq:boostfactor_uni}) can be reduced to $\cos(\gamma)=\frac{1}{n|\Vec{\beta}^*|}$, with $\gamma$ the angle between $\Vec{\beta}^*$ and $\hat{r}$, allowing one to recover the standard expression for the Cherenkov angle.
\subsection{Travel time and directional reconstruction}\label{sec:TravelTime}
\begin{figure}
    \centering
    \includegraphics[width=\linewidth]{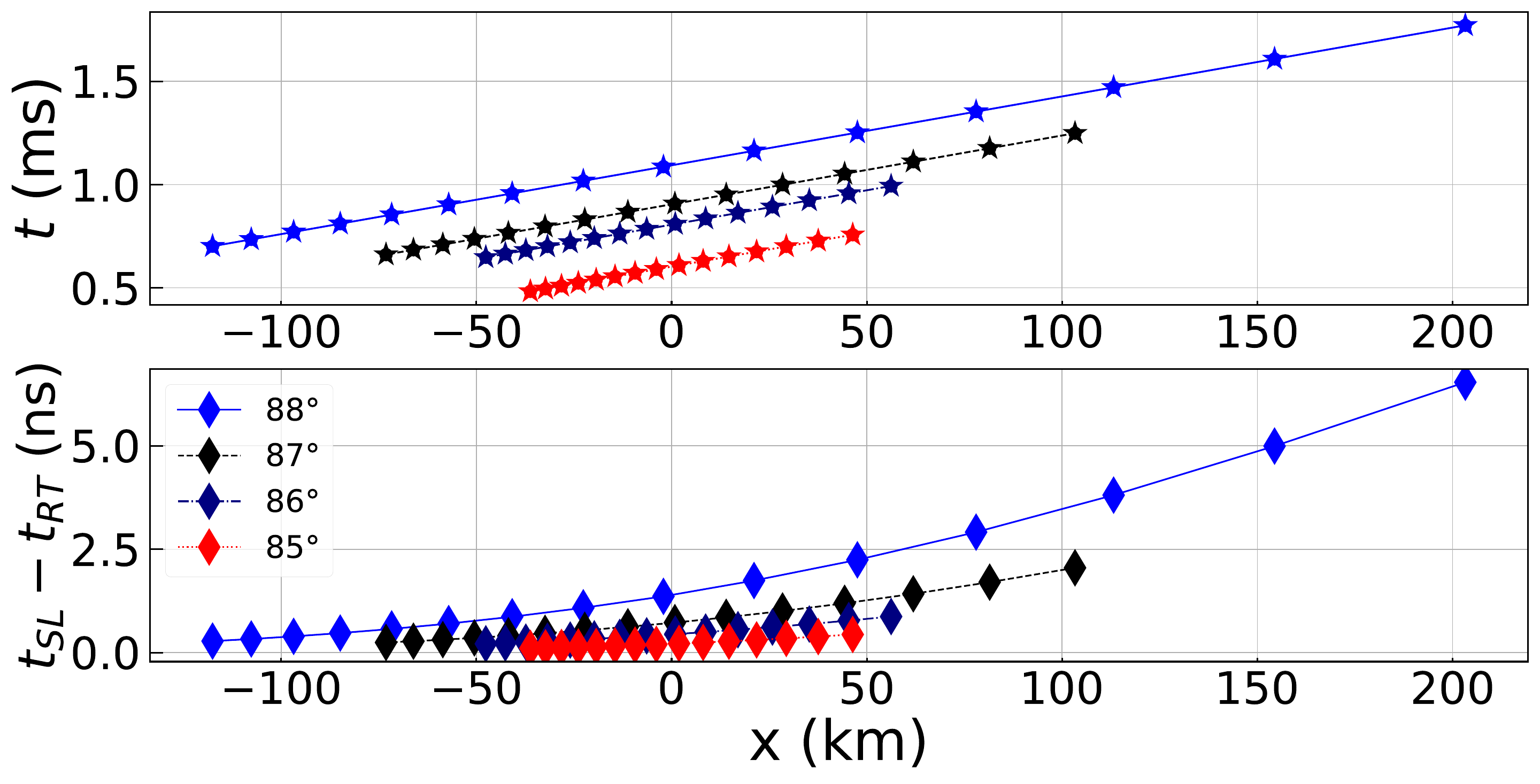}
    \caption{Top: Travel time for different receivers placed at $z=0$ in function of their $x$ coordinate. Rays are traced from an estimated position of $X_\mathrm{max}$ to the z=0 coordinate, the geometry is shown in Fig~\ref{fig:RTParams} and Fig.~\ref{fig:ReconstructionDiagram}. Shown are the travel times for both the straight line approximation(dots) as well as the more correct travel times from raytracing(stars) with both results overlapping. Bottom: The difference between the straight line travel time and the raytraced travel time.
    Note that the range of values on the $x$-axis becomes larger for more inclined showers due to an increased size of the signal footprint. }
    \label{fig:TimingResiduals}
\end{figure}

\begin{figure}
    \centering
    \includegraphics[width=\linewidth]{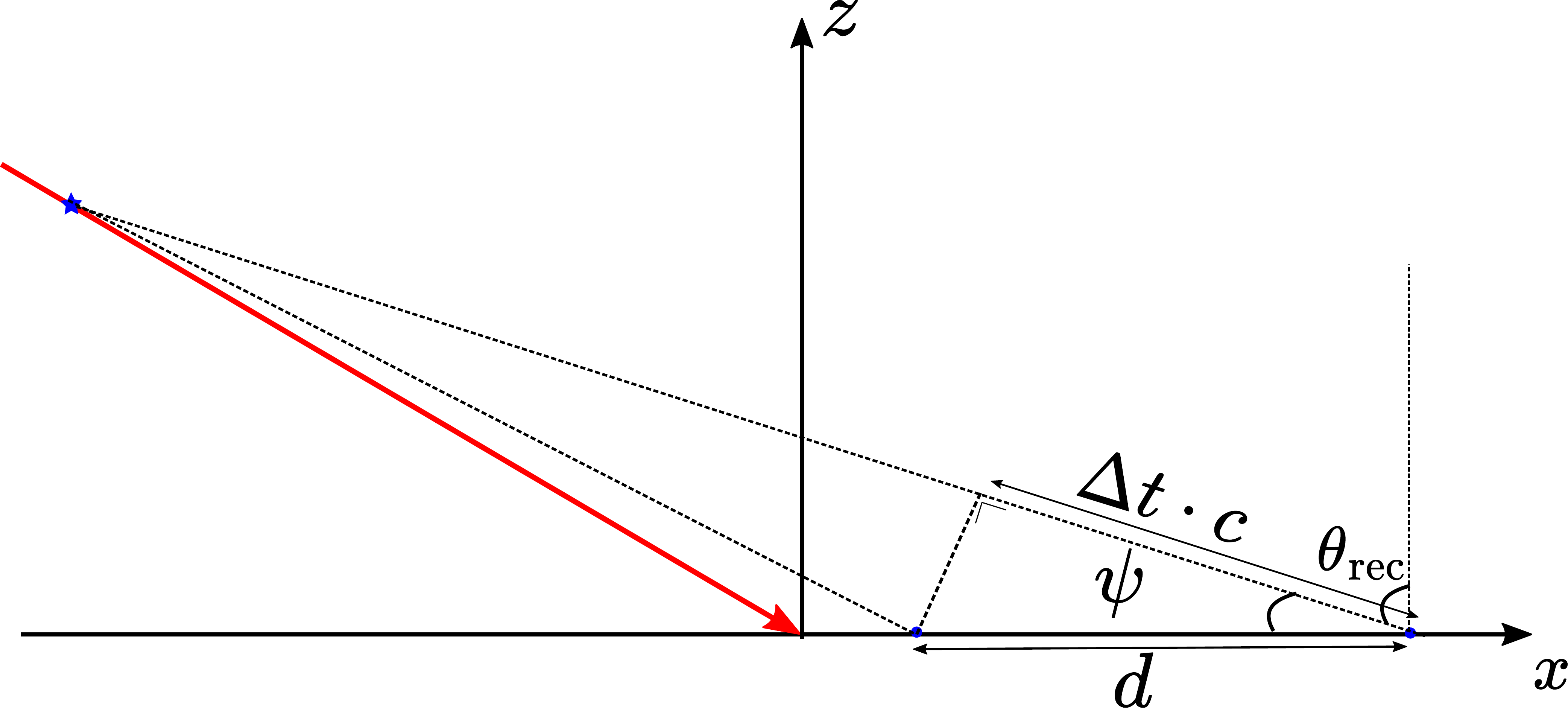}
    \caption{Diagram to explain directional reconstruction. The angle $\psi$ can be reconstructed from the distance $d$ between receivers as well as the difference in signal arrival times $\Delta t$.}
    \label{fig:ReconstructionDiagram}
\end{figure}

By use of the raytracer and the line model of the cascade explained above, one can look into the difference in travel times for a straight line and a curved ray path. 
The straight line travel time $t_{\mathrm{SL}}$ and the curved ray path travel time $t_{\mathrm{RT}}$ can be computed as:
\begin{gather}
    t_{\mathrm{SL}}=\frac{\langle n\rangle}{c}\cdot R
    \\
    t_{\mathrm{RT}}= \frac{L}{c}
\end{gather}
To investigate the effect of ray curvature on travel times we make use of the geometry as shown in Fig.~\ref{fig:ReconstructionDiagram}, where rays are traced from an estimated position of $X_{\mathrm{max}}$ until coordinate $z=0$. For each of these traced rays one can then compare the difference between $t_{\mathrm{SL}}$ and $t_{\mathrm{RT}}$. 
The result of this comparison is shown in Fig.~\ref{fig:TimingResiduals} for a range of different zenith angles. 
\par
The results shown in Fig.~\ref{fig:TimingResiduals} are in agreement with earlier work~\cite{Alvarez-Muniz:2015ayz,Schluter:2020tdz,Werner:2007kh} where it was shown that for zenith angles up to $85^{\circ}$ the difference in travel time is of $O(0.1)$ ns. Do note that the difference in travel times increases for more inclined showers, up to $5$ ns for $\theta=88^{\circ}$ at axis distances of 200 km. These values are much larger than the $0.01$ ns values found for $\theta <85^{\circ}$  shower, but still small compared to the millisecond travel times. Also note that it is not the difference in travel time of a single receiver point, as shown here, but rather the arrival time differences between different antennas that are used in directional reconstruction.
\\
\par
An example of how timing differences can be used for directional reconstruction is shown in Fig.~\ref{fig:ReconstructionDiagram}, 
where the relative arrival times between antennas can be used to determine the angle $\psi$ as 
\begin{equation}
    \cos{\psi}= \frac{\Delta t \cdot c}{d}
\end{equation}

\begin{figure}
    \centering
    \includegraphics[width=\linewidth]{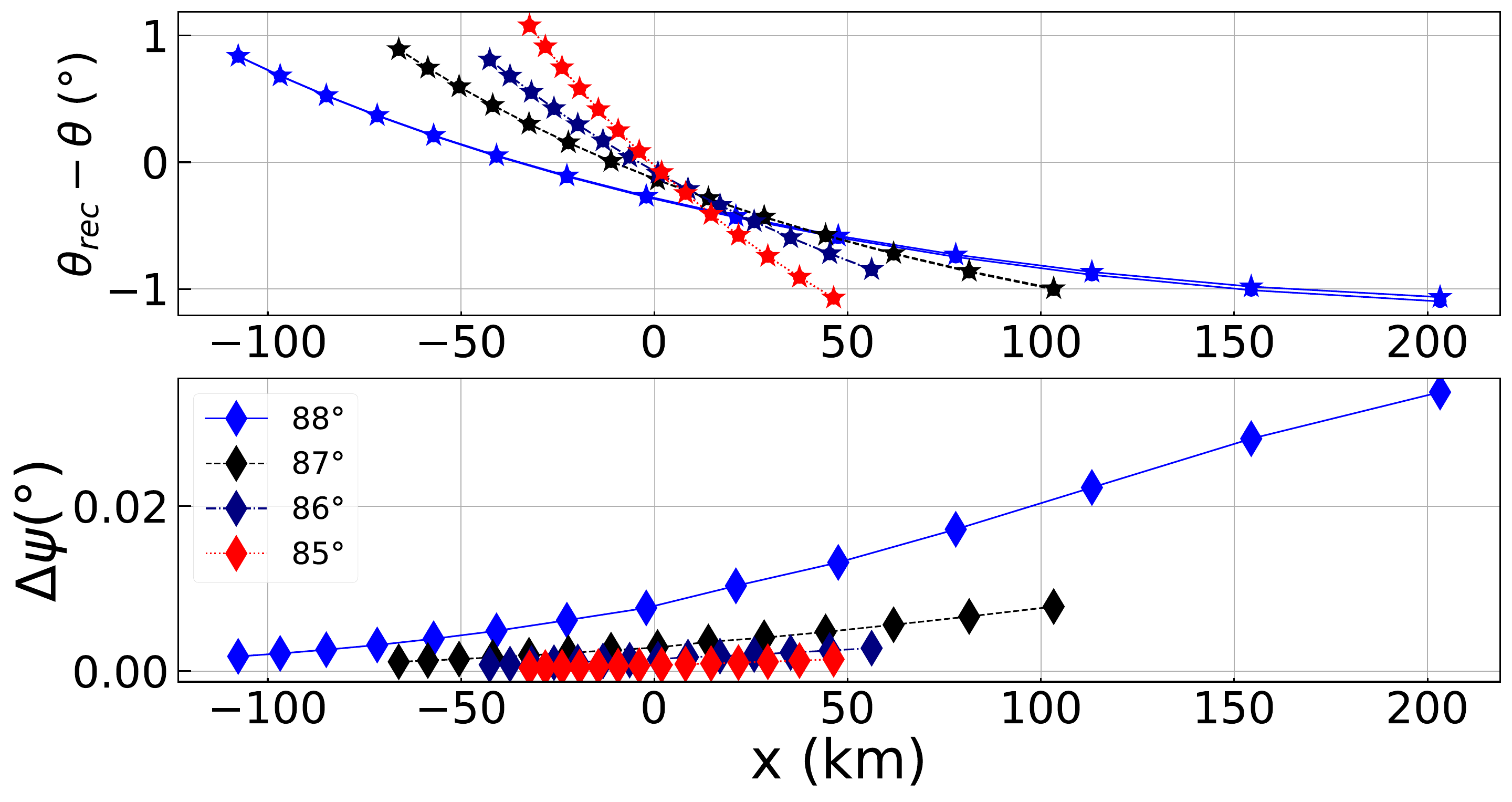}
    \caption{Top: The difference between the zenith angle of the shower $\theta$ and the reconstruction of the zenith angle $\theta_{rec}=90^{\circ} - \psi$ based on the angle $\psi$ computed from timing differences between neighboring receivers, that is the difference in arrival times between a receiver at a coordinate x and the nearest receiver in the direction of negative x. Shown are both the results based on timing differences from a straight line approximation (in dots) and timing differences from the curved ray path travel times (stars), with both results nearly overlapping.
    Bottom: the difference between the angle $\psi$ as calculated by using the timing differences from a straight line approximation minus the angle calculated with the more correct timing differences from ray tracing. This value is an estimate of the uncertainty on directional reconstruction due to a uncertainty in the timing differences, of the order $O (0.01^{\circ})$.
    For the geometry related to this reconstruction, see Fig.~\ref{fig:ReconstructionDiagram}.}
    \label{fig:AngleOffsets}
\end{figure}

Moving to the more realistic curved ray path then would alter the values of $\Delta t$ and in turn give a different value for $\psi$. This allows the estimate of the introduced uncertainty when determining $\psi$ from timing differences assuming straight-line signal propagation. A comparison is shown in Fig.~\ref{fig:AngleOffsets} 
for zenith angles between $85^{\circ}$ and $88^{\circ}$. 
\par
From this we can conclude that the directional reconstruction difference introduced by the straight line approximation should remain at most of the order $O(0.01^{\circ})$ up to zenith angles of $88^{\circ}$, which is well below the accuracy of detection efforts.
 
\subsection{Discussion of the boostfactor}\label{sec:BoostfactorDiscussion}
\begin{figure*}
    \includegraphics[width=0.49\linewidth]{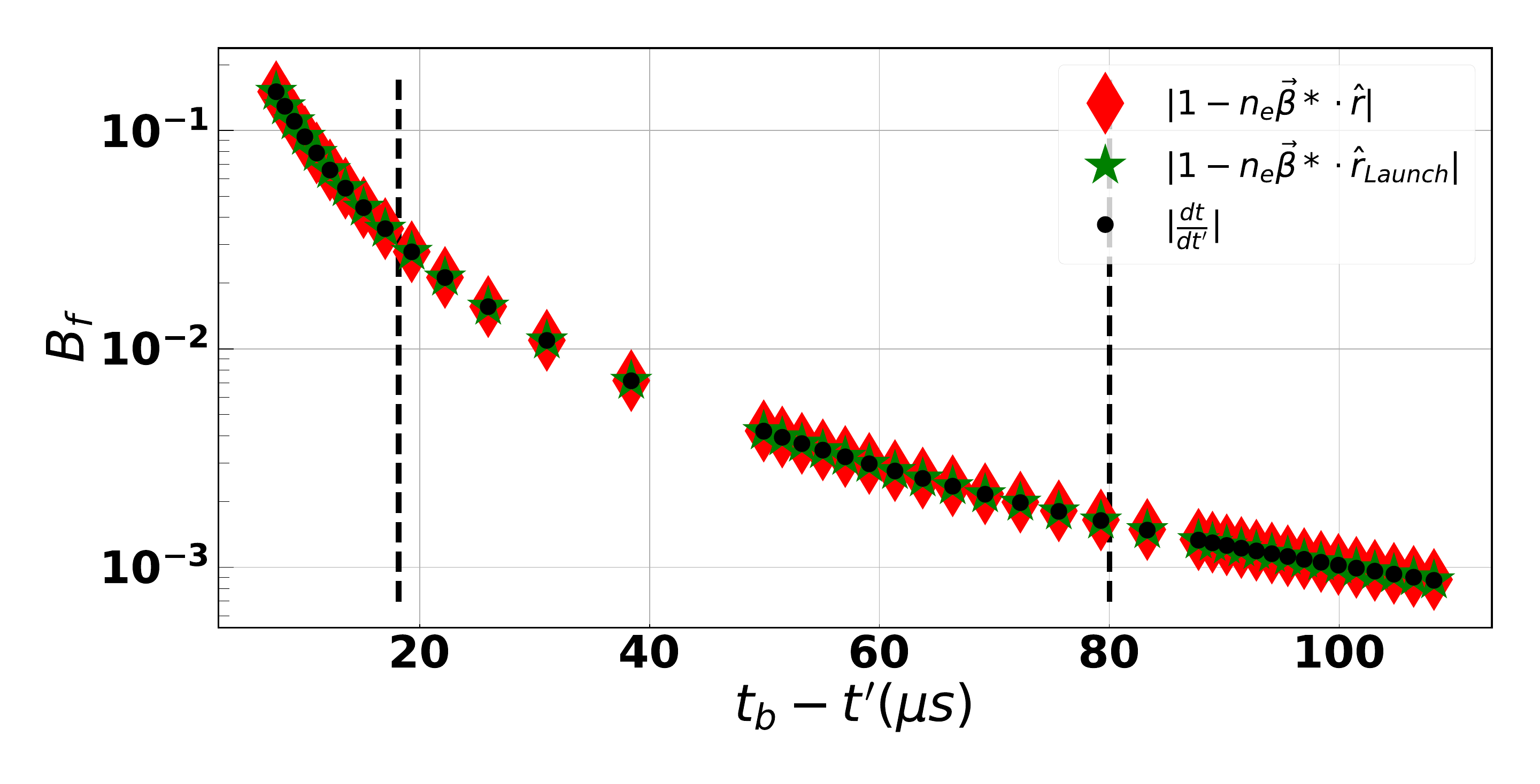}
    \includegraphics[width=0.49\linewidth]{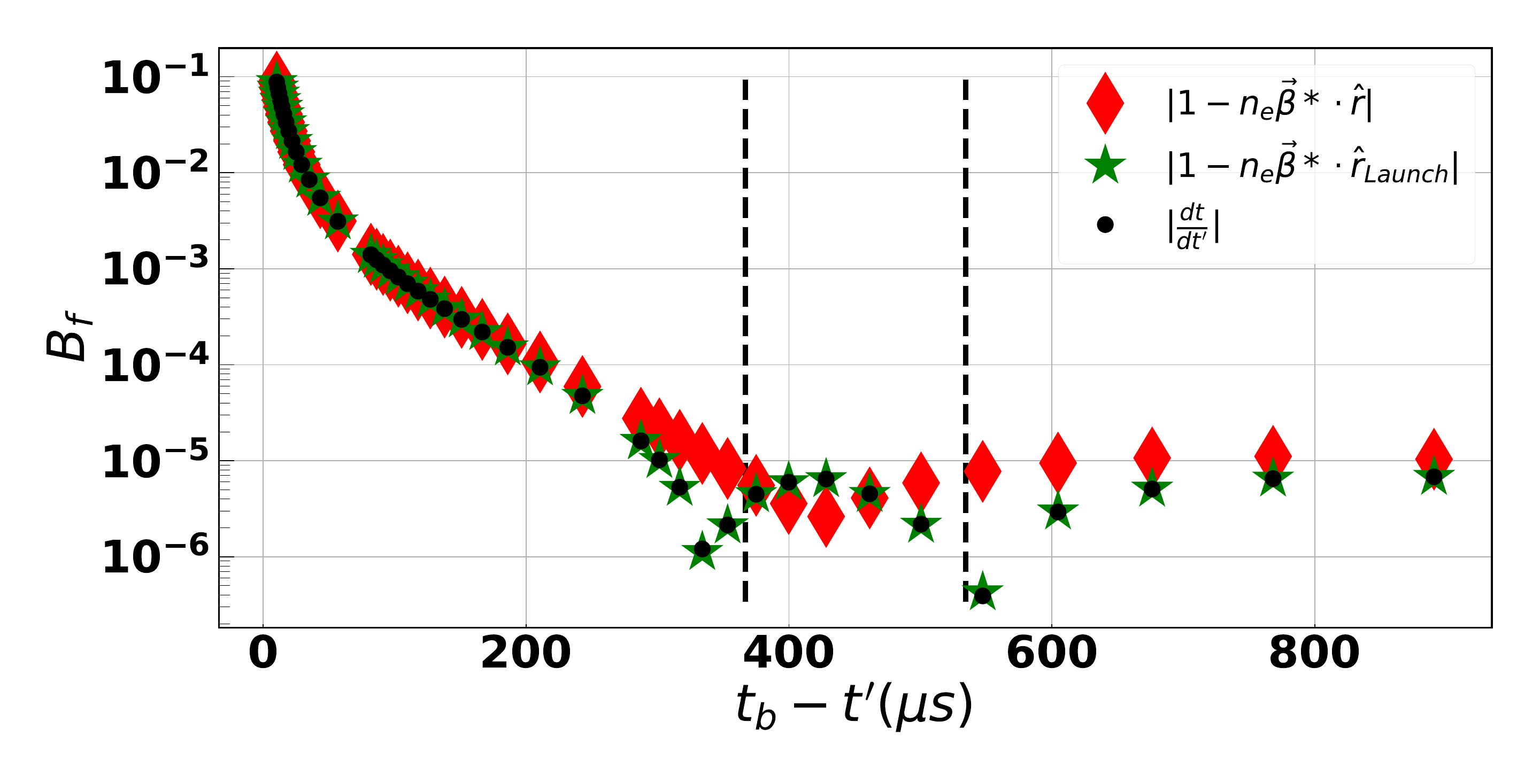}
    \caption{The boostfactor as a function of the difference between time of closest approach and emission times for a zenith angle of $65^{\circ}$ (left) and $85^{\circ}$ (right). The dots (black) correspond to the numerically computed derivative, representing the true value of the boostfactor, while the diamonds (red) and stars (green) correspond to the boostfactor calculation with $\hat{r}$ and $\hat{r}_{\mathrm{Launch}}$ respectively. The area between the vertical black dashed lines denotes a typical region around the point where one would find the maximum amount of emitters ($X_{\mathrm{max}}$) to give an indication on which part of the cascade causes the largest part of the emission. This region is determined from the electron content of the cascade as  shown in Fig.~\ref{fig:LongPlot}.}
    \label{fig:Boostfactor}
\end{figure*}

\begin{figure}
    \centering
    \includegraphics[width=\linewidth]{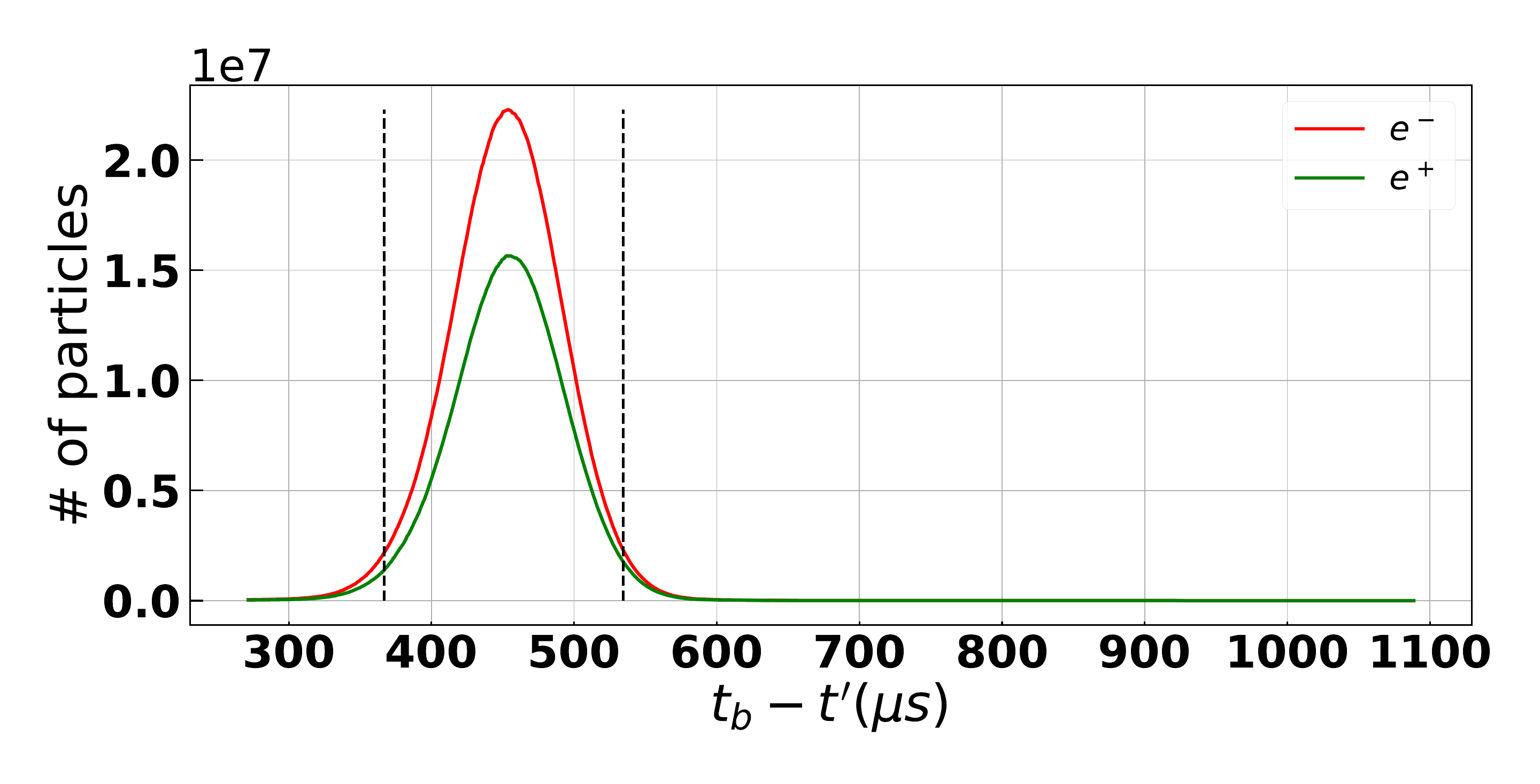}
    \caption{The electron and positron content of the $\theta=85^{\circ}, E=10^{17} eV$ proton primary example shower used to construct the right plot of Fig.~\ref{fig:Boostfactor}. The vertical black dashed lines are placed at the same position as in the right plot of Fig.~\ref{fig:Boostfactor} and mark the region where the electron content is at least 10\% of the peak value, giving an indiction on which part of the cascade causes the largest part of the emission.}
    \label{fig:LongPlot}
\end{figure}

\begin{figure}
    \centering
    \includegraphics[width=\linewidth]{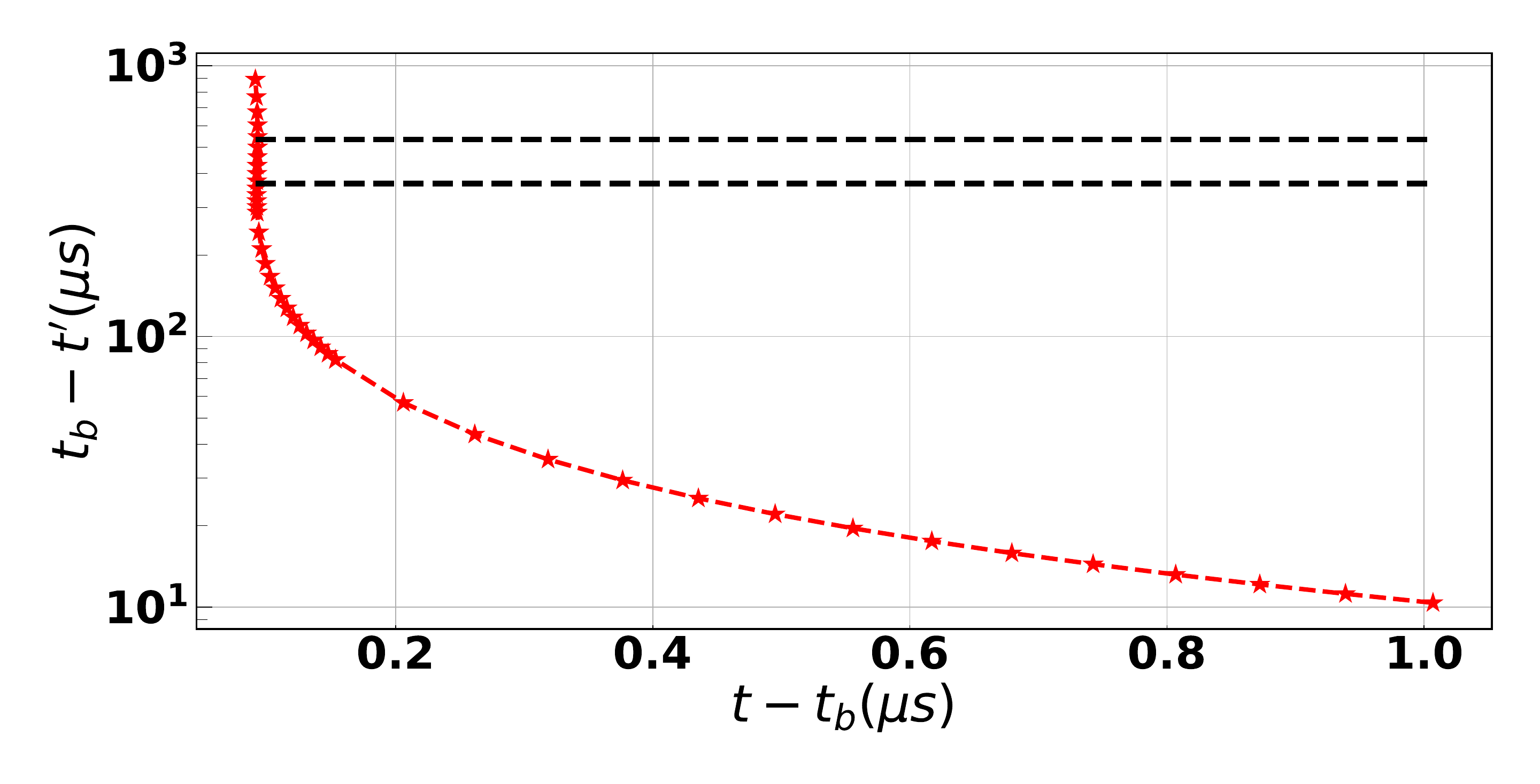}
    \caption{The relation between emission times $t'$ and arrival times $t$ as related to the time of closest approach $t_b$. Note the strongly boosted region where many different emission times  correspond to a near identical arrival time. The black lines denote a region of the cascade which causes the largest part of the emission. This comparison is presented in a form comparable to the results shown in~\cite{Werner:2007kh}.}
    \label{fig:PlotKWOS}
\end{figure}

To investigate the correct way of formulating the boostfactor for nonuniform media, we compared different formulations of the boostfactor $B_f$ to the correct numerically computed value of $\frac{\mathrm{d}t}{\mathrm{d}t'}$. For this comparison the following formulations of the boostfactor were considered:
\begin{gather*}
    B_f =  |1 - n \hat{r} \cdot \vec{\beta^*}|,
    \\
    B_f =  |1 - n \hat{r}_{\mathrm{Launch}} \cdot \vec{\beta^*}|,
\end{gather*}
with $n$ evaluated at the emitter and $\hat{r}$, $\hat{r}_{\mathrm{Launch}}$ defined as shown in Fig.~\ref{fig:RTParams}.
 A visualization of this comparison for the line model of the cascade is given in Fig.~\ref{fig:Boostfactor}, where the boostfactor was calculated as a function of the difference between the time of closest approach between shower front and receiver $t_b$  and emission time $t'$. Results of the ray tracer are shown for receivers placed at a perpendicular distance of $1400$ m from the shower axis.
The electron and positron content of the $85^{\circ}$ shower as a function of $t_b-t'$ is shown in Fig.~\ref{fig:LongPlot}. 
For reference we also show the relation between emission times and arrival times for our $85^{\circ}$ example in Fig~\ref{fig:PlotKWOS}
, made in a comparable manner to the results shown in~\cite{Werner:2007kh}.
Note that the $85^{\circ}$ comparison of Fig.~\ref{fig:Boostfactor} shows two dips, corresponding to two different regions of strong boosting where $\frac{\mathrm{d}t}{\mathrm{d}t'} \approx 0$. While these two distinct regions of strong boosting are an interesting feature of working with an exponential index of refraction profile, in practice only one of the regions will typically be occupied by the emitting part of a particle cascade. This is due to the large difference in geometrical distance between the two regions of strong boosting.  
From Fig.~\ref{fig:Boostfactor} one can see that the calculation of the boostfactor that uses the index of refraction at the emission point and the initial launch direction of the ray agrees with the numerically computed true boostfactor. This is a nontrivial result which, while not analytically proven, has been verified for many different geometries and has already previously been applied to the simulation of radio emission from cascades passing from air into ice as seen for an in-ice receiver~\cite{DeKockere:2024qmc}. 
Figure~\ref{fig:Boostfactor} also shows that the effects for low zenith angles are not visible by eye on the plots, but that the commonly used straight line calculation of the boostfactor can differ more than an order of magnitude from the correct result for very inclined showers. Care should be taken in interpreting these results, as relating the effects of a changing boostfactor to effects in expected received signal is highly nontrivial and is the subject of the following sections.
\\
\par
\begin{figure}
    \centering
    \includegraphics[width=0.8\linewidth]{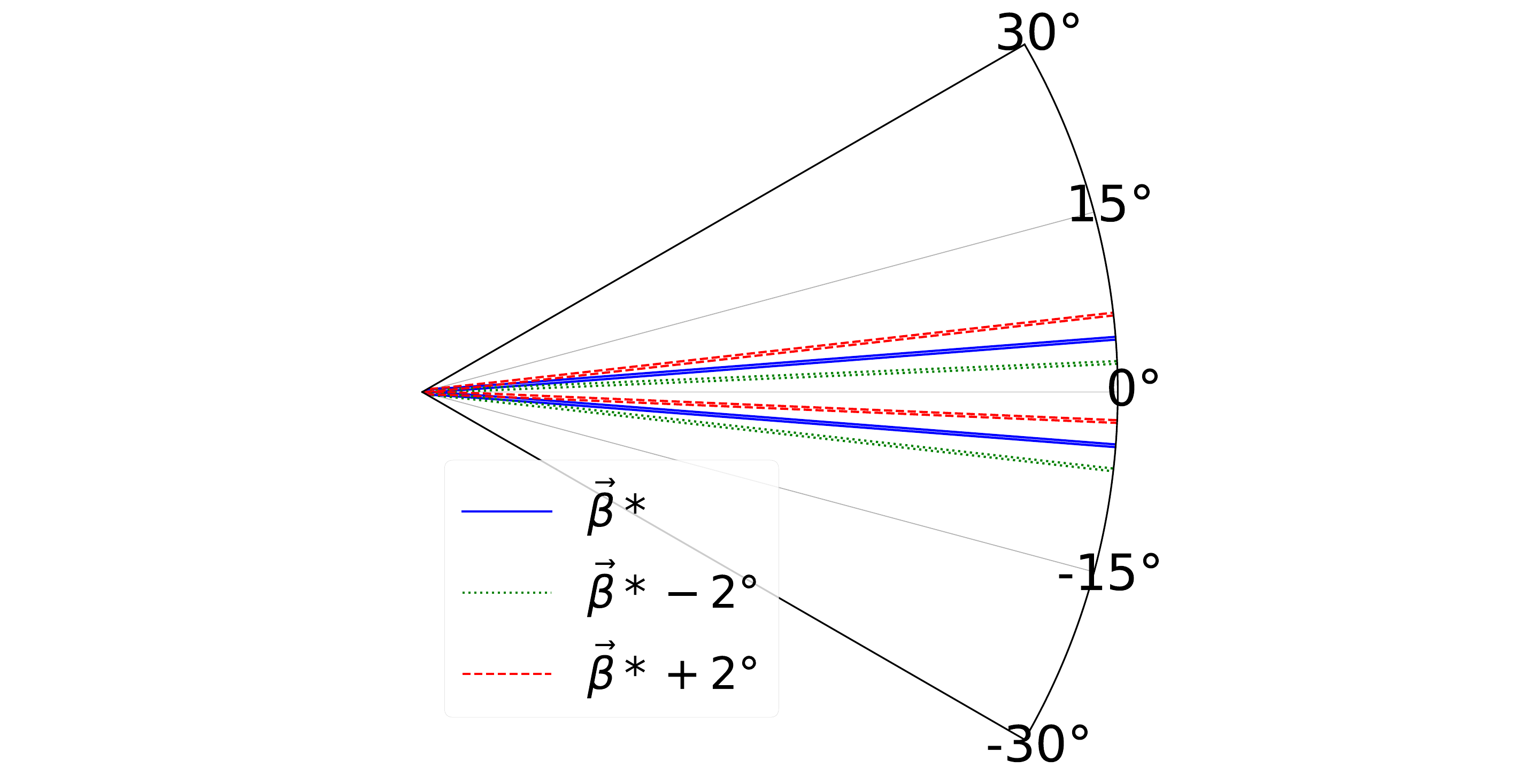}
    \caption{Polar plot showing the radiation pattern based on Eq. \ref{eq:PolarPlot} with $n=1.003$. The blue line represents the radiation pattern which peaks at the Cherenkov angle for a horizontal $\vec{\beta^*}$ propagating along the line $0^{\circ}$. The red and green lines correspond to the radiation pattern where $\vec{\beta^*}$ is shifted up or down $2^{\circ}$. Realistic values are expected to be of the order $1^{\circ}$~\cite{Lafebre:2009en}. A shift in initial direction could lead to a larger expected signal for some subset of emitters and a lower expected signal for some other subset, allowing the effects to potentially cancel out.  }
    \label{fig:PolarPlot}
\end{figure}

Note that the results shown in Fig.~\ref{fig:Boostfactor} represent only the denominator of Eq.(~\ref{eq:End-point}). 
To have an idea of the full effect of moving to a curved ray path, one should also replace the $\hat{r}$ in the numerator of Eq.(~\ref{eq:End-point}) and consider the full term $\left( \frac{\hat{r} \times [\hat{r} \times \vec{\beta}^*]}{|1-n\vec{\beta}^* \cdot \hat{r}|} \right)$. To qualitatively estimate the effect of changing $\hat{r}$ in the full term, consider a region close to an emitter where $n$ can be assumed to be constant. Assume also that $|\vec{\beta}^*|\approx 1$. Defining $\alpha$ as the angle between $\hat{r}$ and $\vec{\beta}^*$ we can then write:
\begin{equation}
     \frac{\hat{r} \times [\hat{r} \times \vec{\beta}^*]}{|1-n\vec{\beta}^* \cdot \hat{r}|} = \frac{\sin{\alpha}}{|1-n \cos \alpha|}. \label{eq:PolarPlot}
\end{equation}
The line model of the cascade introduced earlier assumed that all the emitters had their momentum aligned along the shower axis, in other words along the red dashed line of emitters in Fig.~\ref{fig:RTParams}.
Consider now also that in a realistic air shower geometry, there will be some spread regarding the direction of $\vec{\beta}^*$ around the shower axis. A different value of $\vec{\beta}^*$ would mean a different direction for the Cherenkov cone: the maximum of  the right-hand side of Eq.~\ref{eq:PolarPlot}. The effect of the spread on the maximum is illustrated in Fig.~\ref{fig:PolarPlot}. Note that the typical spread of $\vec{\beta}^*$ would be of order $1^{\circ}$ around the shower axis~\cite{Lafebre:2009en}.
This spread in $\vec{\beta}^*$ could cancel out the effects of a change in initial propagation direction due to ray curvature, as a shift in initial direction would lead to a direction closer to the Cherenkov cone associated with some subset of emitters and further away from the Cherenkov cone associated with some other subset, allowing the effects to potentially cancel.
\\
\par
\begin{figure}
    \centering
    \includegraphics[width=\linewidth]{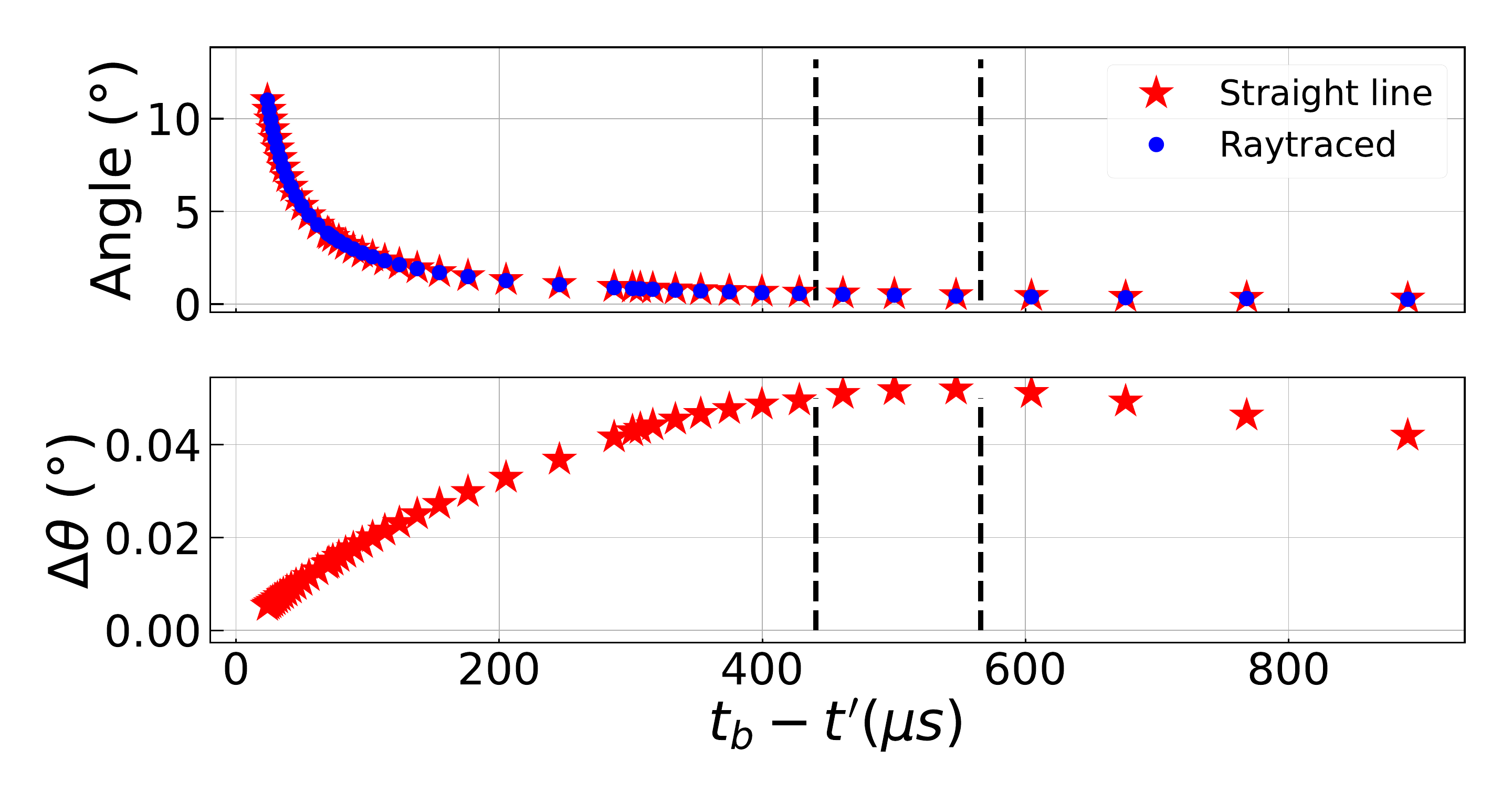}
    \caption{ 
    Top: The angle, in degrees, between the direction of initial propagation and the shower axis for both a ray traced path (blue dots) and a straight-line path (red stars) for a geometry with a zenith angle of $85^{\circ}$. Bottom: The difference between the straight line and ray trace angle , in degrees, shown in the top plot. Note that this difference reaches a  value of $0.05^{\circ}$, much smaller than the typical $1^{\circ}$ of spread for the momentum around the shower axis. The two black lines represent a typical region around $X_{\mathrm{max}}$ as in Fig.~\ref{fig:Boostfactor}. }
    \label{fig:AngleDifference}
\end{figure}

As part of this argument, related to the spread of momentum, one should also investigate the change in initial signal propagation direction induced by moving to a curved ray path. 
The difference in initial propagation direction due to ray curvature can again be quantified in terms of angles: either the straight line angle between $\hat{r}$ and the shower axis or the launch angle between $\hat{r}_{\mathrm{Launch}}$ and the shower axis.
Figure~\ref{fig:AngleDifference} shows the comparison between the two angles for a geometry with a zenith angle of $85^{\circ}$, a similar geometry as used to obtain the results shown in Fig.~\ref{fig:Boostfactor}. Also shown is the difference between the two angles, found to reach a maximum value of $0.05 ^{\circ}$, far smaller than the momentum spread in a realistic air shower. 
Note that while the results in Fig.~\ref{fig:Boostfactor} show large differences between ray traced and straight line trajectories on the boostfactor level, the small absolute difference in angles shown in Fig.~\ref{fig:AngleDifference} gives a more nuanced picture. As explained above: a spread in emitter momentum could potentially smear out the effect of moving to a curved ray path. We also mention that the difference between ray-trace and straight-line angles grew when moving to more inclined geometries, but that they remain smaller than the typical momentum spread in air showers.
\\
\par
Finally, note that the $R$ present in Eq~\ref{eq:End-point} is the geometrical distance between emitter and receiver. To verify that it should be the geometrical distance $R$ and not the optical path length $L$ one can look at the example shown in Appendix~\ref{sec.appendixD}, where we calculate the electric field for a medium with constant $n > 1$ and recover the standard Coulomb field in the case of a stationary charge.
\\
\par
To investigate ray curvature effects on the simulated radio signal in a more complete manner, we use an adapted version of CoREAS which is able to modify the end-point calculation by use of tabulated data from the ray tracer. An explanation of this tabulation procedure is given in appendix~\ref{sec.appendixA}. Note that we assume all emitters to be on the shower axis for the calculation of the boostfactor, this is expected to be a reasonable assumption due to the distances between emitter and shower axis being small compared to the distances between emitter and receiver. A more in-depth verification of this assumption is given in Appendix~\ref{sec.appendixB}. For this paper, we used version 7.7420 of the CORSIKA software.
\section{Simulation setup and geometry}
For the simulation geometry we chose the configuration as shown in Fig.~\ref{fig:CorsikaGeom}, where receivers were placed on a curved line of constant altitude following Earth's curvature in a plane that fully contains the shower axis, defining the origin as the point where the shower axis crosses sea level. We then generate a shower with a specified zenith angle $\theta$ and a set azimuthal angle $\phi=0$. This layout of receivers includes the positions furthest away from the cascade for which the ray curvature is most pronounced and for which the effect of the changing boostfactor should be the largest.
\begin{figure}
    \centering
    \begin{tikzpicture}[x=0.3cm,y=0.3cm,z=0.2cm]
    \tikzstyle{every node}=[font=\large]
    \draw[->] (xyz cs:x=-13.5) -- (xyz cs:x=13.5) node[below] {$\mathbf{North}$};
    \draw[->] (xyz cs:y=-5.5) -- (xyz cs:y=10.5) node[right] {$\mathbf{Z}$};
    \draw (0,2)  node[above left] {$\theta$} arc (90:115:5);    
    \draw[ultra thick,->,color=red] (xyz cs:y=10, x=-13.5 ) node[right] {$\mathbf{Shower \ axis}$} -> (xyz cs:y=0,x=0) ;
    \filldraw[color=blue!60, fill=blue!5, very thick] (xyz cs:x=-10,y=1 - 0.28253) circle (0.5);
    \filldraw[color=blue!60, fill=blue!5, very thick] (xyz cs:x=-7,y=1 - 0.12557 ) circle (0.5);
    \filldraw[color=blue!60, fill=blue!5, very thick] (xyz cs:x=-4,y=1 -0.03139) circle (0.5);
    \filldraw[color=blue!60, fill=blue!5, very thick] (xyz cs:x=10,y=1 -0.28253) circle (0.5);
    \filldraw[color=blue!60, fill=blue!5, very thick] (xyz cs:x=7,y=1 - 0.12557) circle (0.5);
    \filldraw[color=blue!60, fill=blue!5, very thick] (xyz cs:x=4,y=1- 0.03139) circle (0.5);   
    \draw[color=black!60, very thick] (xyz cs:x=0) circle (0.5) node[below 
    right] {$\mathbf{West}$};
    \draw [dashed] (-13.5,1) -- (13.5,1);
    \draw[color=black!100, very thick] (0,0) node {x};
    \end{tikzpicture} 
    \caption{ Geometry used for CORSIKA/CoREAS simulations. Receivers (blue dots) are placed in the $\mathbf{Z},\mathbf{North}$ plane on a constant altitude. The coordinate origin is defined as in CORSIKA. Note that the receivers are placed on a constant altitude trajectory following the curvature of the earth, and thus move away from the dashed line representing a constant z coordinate. }
    \label{fig:CorsikaGeom}
\end{figure}
 The ray tracer is then used to generate a table for each receiver which allows the calculation of the correct launch vector $\hat{r}_{\mathrm{Launch}}$ from the straight-line vector $\hat{r}$ used during the calculation in standard CoREAS. A further explanation of the tabulation method and the mapping of $\hat{r}$ to $\hat{r}_{\mathrm{Launch}}$ is given in Appendix~\ref{sec.appendixA}. Through this tabulation, one can now compute the adapted $\hat{r}_{\mathrm{Launch}}$ without the computationally expensive requirement of having to ray trace to each individual emitter during simulation. 
 \\
 \par
 For reference, the CoREAS simulations were set up with the following options:
 \begin{itemize}
     \item The high energy hadronic interaction model used was QGSJETII-04 and the low energy hadronic interaction model was URQMD1.3CR
     \item The primary chosen was always a proton with energy $10^{17}$ eV.
     \item The thinning arguments, used to bundle together low energy particles, were put to THIN: [$10^{-6} \ 10^2 \ 0$]. The first value represents the fraction of the primary energy below which thinning is used. The second value gives a maximum weight until when thinning is used. The last value is used for radial thinning, which is turned off for a value of 0.   
     \item The hadronic thinning was put to THINH: [$1 \ 10^2$]. The values here are used to define a difference in thinning fraction and weights relative to electromagnetic thinning.
     \item The energy cuts applied through ECUTS for hadrons, muons, electrons, and photons were put respectively at: [$0.3 \ 0.3 \ 4.01\cdot 10^{-4} \ 4.01 \cdot 10^{-4}$] GeV.
     \item The time resolution for the CoREAS simulation was put at $0.2$ ns.
 \end{itemize}
 For the full discussion of these input parameters, we refer to the documentation available online~\cite{CorsikaWebsite}.

\section{Effect of the boostfactor on received fluence and radiation energy}
 From the time traces obtained through CoREAS one can compute the received fluence $f$ from a set of $N$ samples in time as~\cite{PierreAuger:2015hbf}:
\begin{equation}
    f=\epsilon_0 c \Delta T \sum_{i=0}^{N-1} E^2(t_i),
\end{equation} 
with $\Delta T$ the width of each time bin, $\epsilon_0$ the vacuum permittivity and with $E$ the modulus of the electric field.
To now investigate the effect of the adapted boostfactor calculation we look at the effect on the received fluence. We do this by comparing the fluence $f_{\mathrm{RT}}$ from a time trace where the calculation was done with the launch vector $\hat{r}_{\mathrm{Launch}}$ to the fluence $f$ from a time trace where the calculation was done with the connecting straight-line vector $\hat{r}$, as is done in standard CoREAS. This comparison is done for a range of receiver positions and is repeated for different values of the zenith angle $\theta$ with pulses filtered to desired frequency ranges to get an understanding of what effect including a curved ray path has on the expected received fluence at each receiver position. All showers were induced by a proton primary with energy $10^{17}$ eV and the magnetic field oriented fully in the $z$ direction with a strength of $50 \ \mu$T.
\\
\par
\begin{figure*}
\begin{center}
 \includegraphics[width=0.49\linewidth]{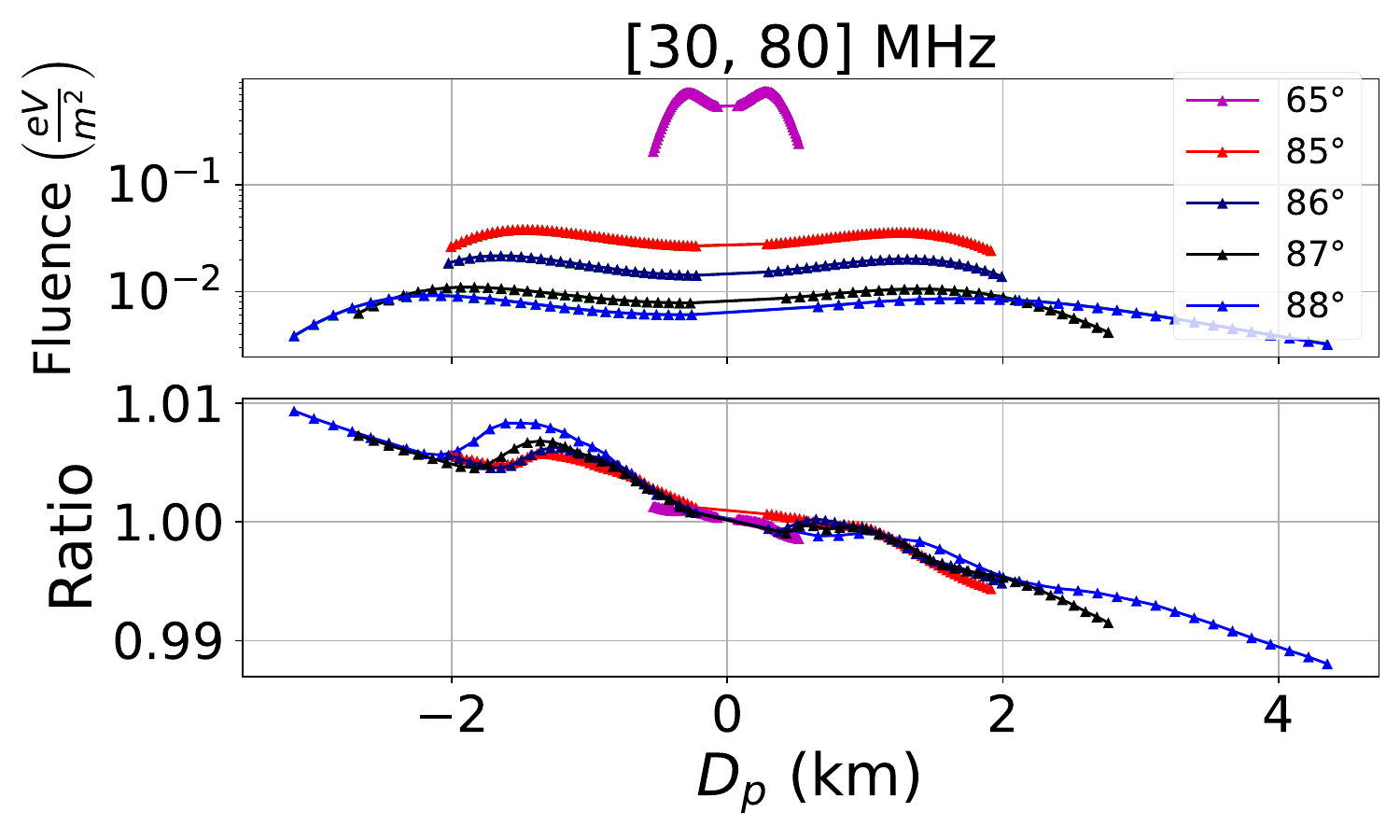}
 \includegraphics[width=0.49\linewidth]{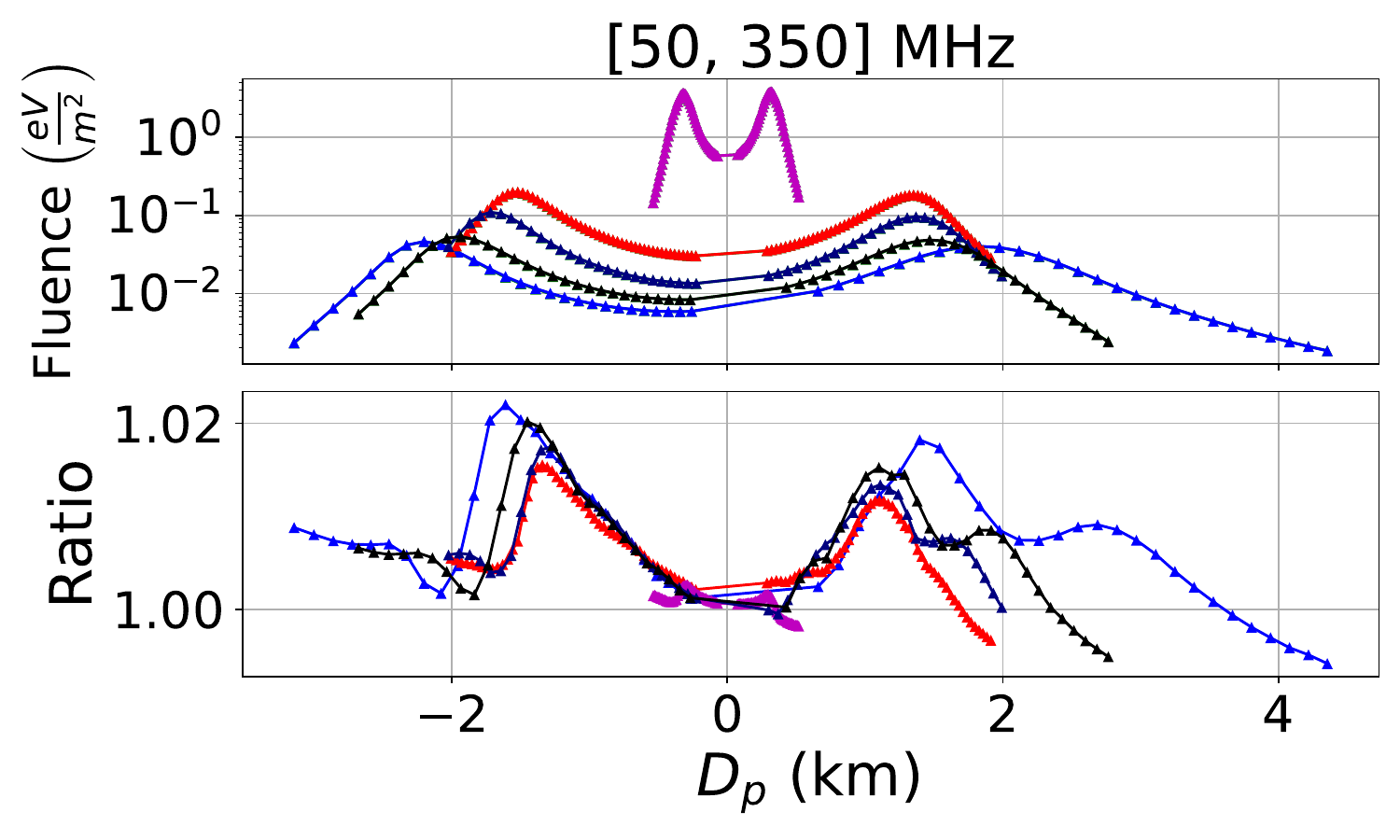}
 \\
 \includegraphics[width=0.49\linewidth]{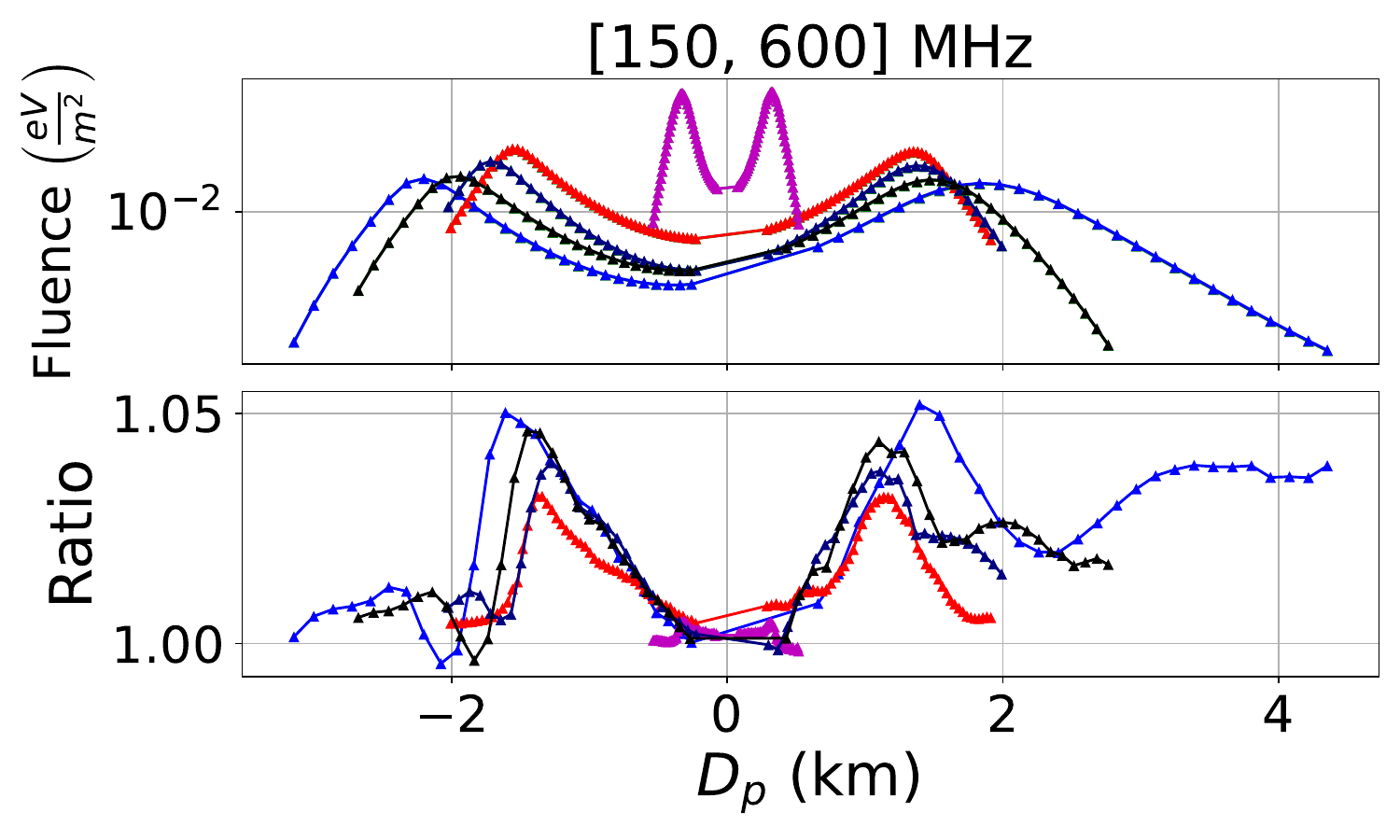}
 \includegraphics[width=0.49\linewidth]{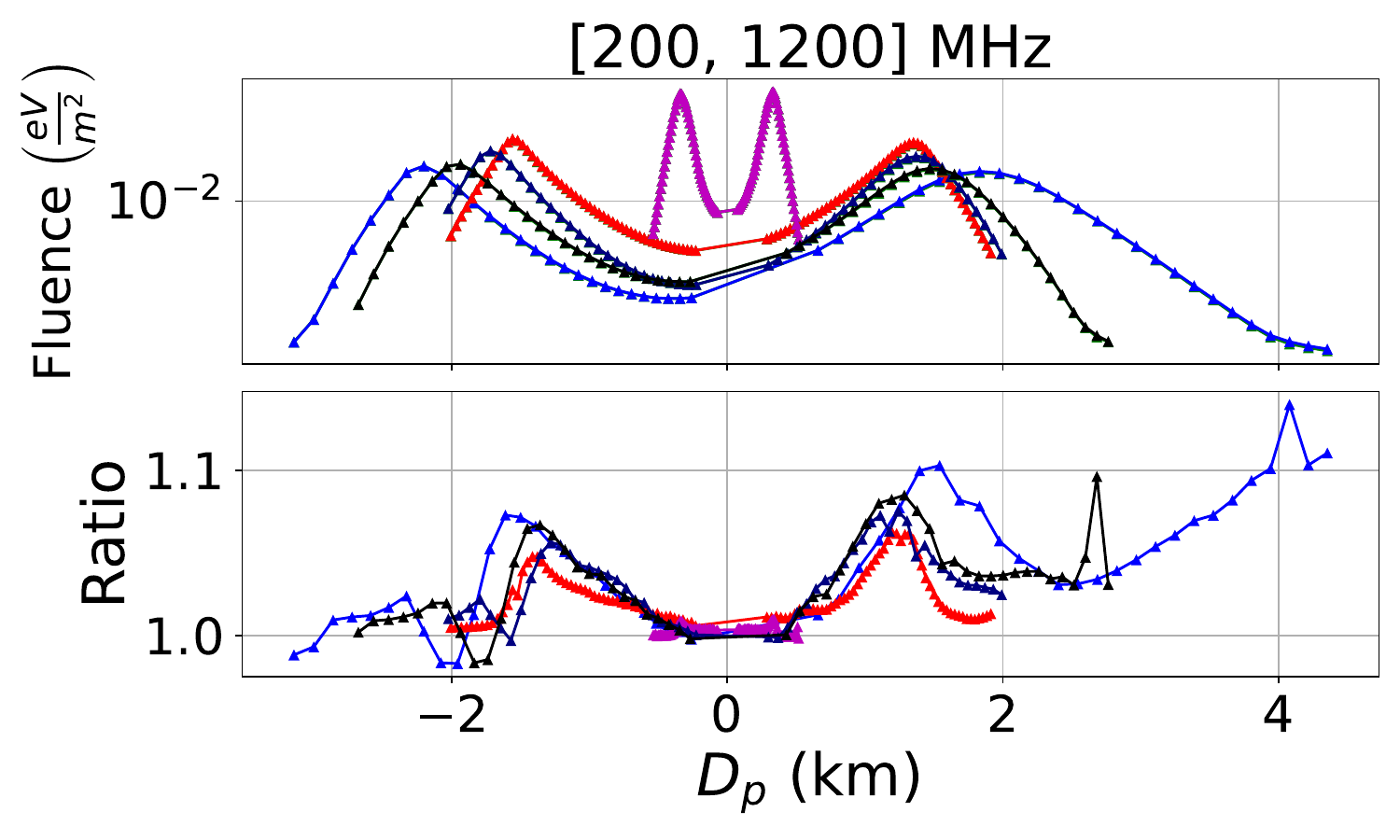}
\end{center}
\caption{Simulated fluences for different frequency ranges for showers with zenith angles of $65^{\circ}$ (magenta), $85^{\circ}$ (red), $86^{\circ}$ ( navy blue), $87^{\circ}$ (black), $88^{\circ}$ (blue). Four frequency ranges are considered, where the top plot always shows the fluence as calculated from CoREAS traces. The lower plot shows the ratio $\frac{f_{RT}}{f}$ between the fluence calculated by using ray tracing data $f_{RT}$ and the fluence $f$ as calculated with an unaltered version of CoREAS. On the x-axis is the distance between the shower axis and the projected position of the observer in the shower plane $D_p$, with negative values corresponding to early geometries and positive values to late geometries. The definition of the projected distance to the shower axis $D_p$ is shown in Fig.~\ref{fig:Early-Late} and its calculation is detailed in appendix~\ref{sec.appendixC}. }
\label{fig:FluencePlots}

\end{figure*}

The fluence comparison for four different frequency ranges and five different zenith angles are given in Fig.~\ref{fig:FluencePlots}, where we show the fluences calculated at different receiver positions as well as the ratio $\frac{f_{\mathrm{RT}}}{f}$ as a function of distance from the shower axis as measured from the projected positions of the observers in the shower plane. The fluence values here have been adapted to correct for early-late effects following the procedure as detailed in~\cite{Schluter:2022mhq}, with the important steps explained in Appendix~\ref{sec.appendixC}. Note that there is a region around the shower axis where no receivers were placed, this is due to technical reasons related to finding a suitable set of initial ray tracing conditions to construct the ray tracing table. 
One can see that the relative difference remains small for the less inclined shower of $65^{\circ}$, with a difference of order $0.1\%$. The difference is highest for the $88^{\circ}$, with some receiver positions showing a 10\% difference. For all other frequencies and zenith angles the values are of single percent level for all receiver positions. This already points toward negligible effects for frequencies up to $600$ MHz and zenith angles up to $88^{\circ}$.
\\
\par
The results in Fig.~\ref{fig:FluencePlots} show the fluence at single receiver points. In the extreme case of radiation up to 1200 MHz for zenith angles of 88° our results indicate fluence deviations of order 10 \%, which would not be negligible. However, to determine the influence on the reconstruction of the energy of a primary particle, it is more instructive to look at the \emph{radiation energy}, i.e, the fluence integrated over the area of the complete radio-emission footprint. We estimate the difference in radiation energy arising from the ray-traced and straight-line signal propagation implementations by treating the ``late'' and ``early'' halves of the radio-emission profiles shown in Fig.~\ref{fig:FluencePlots} separately, assuming circular symmetry for each of the two and then taking the mean of the late and early ratios between the radiation energy calculated with $f$ values and the radiation energy calculated with $f_{\mathrm{RT}}$ values. The relative difference between the two radiation energy calculations for $\theta=65^{\circ}$ was found to be below $1\%$ across all frequency ranges. The relative difference grows when moving to higher frequencies up to a maximum of $3.5\%$  for emission in the 200-1200 MHz band and zenith angles of $88^{\circ}$. For each frequency range considered here, the integrated fluence ratio $\mathcal F$ changes little for different zenith angles between $85^{\circ}$ and $88^{\circ}$ and the average relative difference $(\langle \mathcal{F} \rangle-1)\cdot 100$ is shown in table~\ref{tab:AvgErrFluence} for all considered frequency ranges.
\begin{table}[h!]
\centering
\begin{tabular}{|c|c|}
\hline
\textbf{} & \textbf{$(\langle \mathcal{F}\rangle-1)\cdot 100$}  \\ \hline
\textbf{[30,80] MHz} & 0.09 \%  \\ \hline
\textbf{[50,350] MHz} & 0.8 \%  \\ \hline
\textbf{[150,600] MHz} & 1.9 \%  \\ \hline
\textbf{[200,1200] MHz} & 3.5 \%  \\ \hline
\end{tabular}
\caption{The average uncertainty on radiation energy introduced by straight-line signal propagation for different frequency ranges as calculated for zenith angles between $85^{\circ} - 88^{\circ}$.}
\label{tab:AvgErrFluence}
\end{table}
\par
From this result, we conclude that the effect of the straight line approximation gives a difference of a maximum value of $3.5 \%$ for frequencies up to $1.2$ GHz and zenith angles up to 88$^{\circ}$, with the uncertainty being much smaller, and thus negligible, for less inclined showers or lower frequencies.   
\\
\par
An explanation for the relative difference between fluences, and consequently radiation energies, remaining small while the error on the boostfactor, cf. Fig.~\ref{fig:Boostfactor}, being an order of magnitude can be found in the  implementation of travel times in CoREAS, which takes into account the average index of refraction along the straight line leading to a small uncertainty in travel time of at most $O$ (ns). Due to this, the coherence between different emitters is conserved, leading to a correct result for the bulk emission even though the amplitude for each individual emitter might have an associated uncertainty. This reasoning lies along the same line as the explanation found in~\cite{Alvarez-Muniz:2015ayz,Schluter:2020tdz}, namely that a straight-line path that takes into account the index of refraction profile only introduces an uncertainty of the order of $0.1$ ns for travel times from different parts of the shower and that this small difference does not introduce a significant effect at typically used radio frequencies. Aside from this timing-based argument, the spread in momentum is also expected to cause a cancellation of effects as was explained in section~\ref{sec:TravelTime}.
\\
\par
Care should be taken when applying these results to macroscopic models such as MGMR3D~\cite{Scholten:2019kxcMGMR} and EVA~\cite{deVries:2013bonEVA}, as there the contributions of different emitting elements are typically more significant. One should also note that no Earth skimming effects were taken into account for this study, and that in general geometries where interaction between the radiation and ground become important were not considered here.
\section{Summary and conclusion}
In this paper we explained the concept of the geometric boostfactor and the implementation in CoREAS through the end-point formalism. We showed that the correct way of calculating the boostfactor is by use of the initial launch direction of the ray and the index of refraction value at the emitter, which works in any medium with arbitrarily complex index of refraction profiles. We implemented this adaption of the boostfactor calculation in CoREAS and evaluated the received fluence for a collection of receivers placed on a line as well as a comparison of radiation energies arising from an area integration of the fluence. By comparing the results obtained with the ray tracing data to the results obtained with the standard implementation we found that the relative difference of radiation energies remained at the level of better than 2\%, and thus normally negligible, for zenith angles of up to $88^{\circ}$ and frequencies up to 600 MHz, and reaches a level of $3.5$\% for frequencies of $200-1200$ MHz and zenith angles of $88^{\circ}$. This result is interesting in light of the differences in boost factors arising from curved-ray versus straight-ray calculations reaching factors of up to an order of magnitude. This apparent contradiction can be resolved by two arguments:
\begin{itemize}
    \item The approach to calculating travel time implemented in CoREAS by use of a straight line and an average index of refraction leads to uncertainties of at most $O(1)$ ns, as was discussed in Sec.~\ref{sec:TravelTime}. 
    Because of this, the coherence between different emitters is correctly handled, which is more important than the amplitude of a single emitter. This is also why refractive effects already appear in standard CoREAS as shown in~\cite{Schluter:2020tdz}.
    \item The large error for the boostfactor that was shown in Fig.~\ref{fig:Boostfactor} 
    was made assuming the momentum of each emitter was aligned with the shower axis. In a realistic air shower, however, the momenta of the individual emitters are spread following a relatively broad distribution. As was explained in Sec.~\ref{sec:BoostfactorDiscussion}, 
    such a spread would serve to even out the error over an entire population of emitters, as some subset of emitters would experience stronger boosting and some others would experience weaker boosting when the initial direction of propagation is changed slightly by moving to a curved ray-path. The typical spread of momentum around the shower axis is of the order $O(1^{\circ})$ while the shift in initial direction due to a curved ray path is of the order $O(0.01^{\circ})$.
\end{itemize}
We conclude that, for frequencies up to 600 MHz and zenith angles up to $88^{\circ}$, the effect of moving to a curved ray path remains negligible for most applications. The largest difference was found at the most inclined showers and at the highest frequencies, with a $3.5 \%$ error on the radiation energy.
By use of the raytracer we also found an uncertainty of $O(0.01^{\circ})$ when calculating the arrival direction of emission by use of differences in arrival times, which is far below the accuracy of detectors. We thus conclude that for all foreseen applications, up to $600$ MHz and $88^{\circ}$ zenith angle, the straight-line approximation remains valid. However, one should take care when assuming a straight-line approximation for geometries with a zenith angle of $88^{\circ}$ or higher when working at GHz frequencies. 
\section{Acknowledgments}
This work has been supported by the the European Research Council under the EU-ropean
Unions Horizon 2020 research and innovation programme. (grant agreement No 805486).
\\
This work has been supported by the Flemish Foundation for Scientific Research. (FWO-G085820N).
\appendix
\section{Tabulation of ray tracing data}\label{sec.appendixA}
This appendix details the construction of the ray tracing tables that are used to improve the accuracy of the boostfactor calculation during a CoREAS simulation. A visual representation of the important parameters for this discussion can be found in Fig. ~\ref{fig:RTParams}.
\\
\par
The tables are generated using the ray tracer specified in the third section. For a specified zenith angle and receiver positions, rays are back-tracked until they hit the shower axis. At the point B where a ray hits the shower axis the x components of both the launch vector and the unit vector along the straight-line between point B and the receiver are saved to a table. This is done for a dense sampling of the shower axis, allowing one to get a relation between the $x$-components of the straight-line vector $\hat{r}$ and the $x$-component of the launch vector $\hat{r}_{\mathrm{Launch}}$. 
Note that due to the way these tables are constructed we implicitly assume that the emitter is placed at the shower axis. This is acceptable as we expect the distances between shower-axis and emitter to be small relative to the typical distances between emitter and receiver.
\\
\par
\begin{figure}
    \centering
    \includegraphics[width=\linewidth]{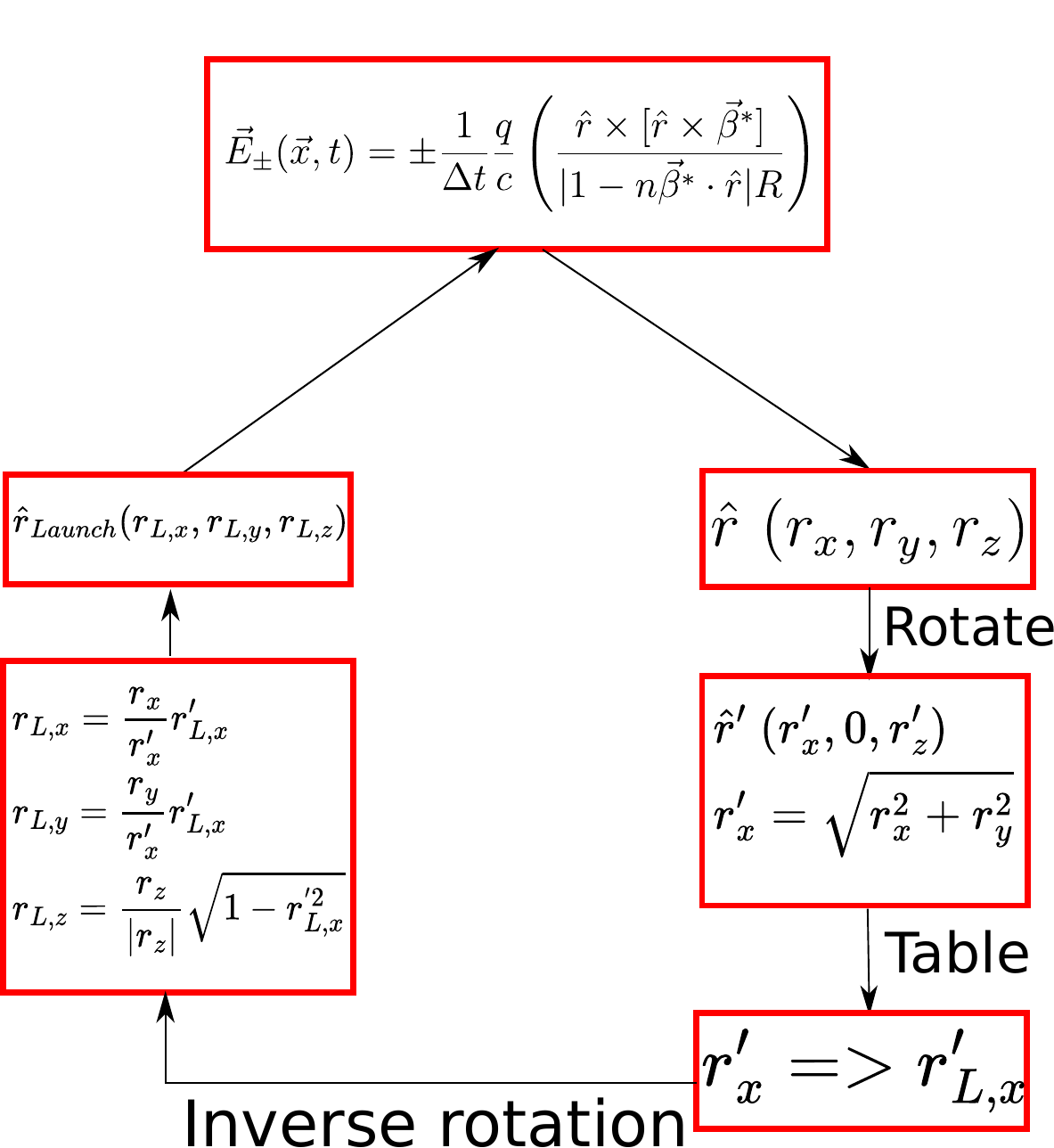}
    \caption{ Diagram explaining the usage of tables during a CoREAS simulation, starting from the expression of the end point formalism and ending with the construction of the launch vector $\hat{r}_{\mathrm{Launch}}$ to replace the straight-line vector $\hat{r}.$ }
    \label{fig:Diagram}
\end{figure}

An explanation of how the tables are used is given below, with a diagram explaining the main steps given in Fig.~\ref{fig:Diagram}.
To explain how the tables are used first note that the ray tracer works with a 2D coordinate system, while CoREAS works with a 3D geometry. We therefore rely on spherical symmetry to reduce the 3D geometry in CoREAS to a 2D geometry where we can use the results from the ray tracer. 
\\
\par
Our approach is as follows: assume an emitter and associated unit vector $\hat{r}=r_x \hat{x} +r_y \hat{y} +r_z \hat{z}$ pointing toward the receiver from the emitter along a straight-line. Due to spherical symmetry we can perform a rotation so that we can describe $\hat{r}$ using only the vertical and horizontal components: 
\begin{gather}
    \hat{r}=r_x' \hat{x}' + r_z \hat{z},
    \\
    r_x'=\sqrt{r_x^2 + r_y^2}.
\end{gather}
This $r_x'$ of the straight-line vector is now used with the table to obtain the corresponding value $r'_{L,x}$ of the launch vector from ray tracing. We then perform the inverse rotation back to the original frame, leading to:
\begin{gather}
    r_{L,x}= \frac{r_x}{\sqrt{r_x^2 + r_y^2}} r_{L,x}',
    \\
    r_{L,y}= \frac{r_y}{\sqrt{r_x^2 + r_y^2}} r_{L,x}',
    \\
    r_{L,z}= \frac{r_z}{|r_z|} \sqrt{1-r_{L,x}^{'2}},
\end{gather}
where we obtained $r_{L,z}$ from  $r'_{L,x}$ and the fact that the vector must have norm 1.
Note that we also defined the sign of $r_{L,z}$ to be equal to the sign of the $z$ component of the original straight-line vector $\hat{r}$. In practice, this means that we restrict ourselves to direct rays.
\section{Verification of line model} \label{sec.appendixB}
For the results in this paper, the launch vector was calculated from tabulated data for which a simple line model of the cascade was used. This means that changes in the launch vector due to lateral displacement from the shower axis are not taken into account. This approach is expected to be valid due to the relatively small distances for emitters from the shower axis with respect to the distance between the emitter and receiver.
\\
\par
\begin{figure}
    \centering
    \includegraphics[width=\linewidth]{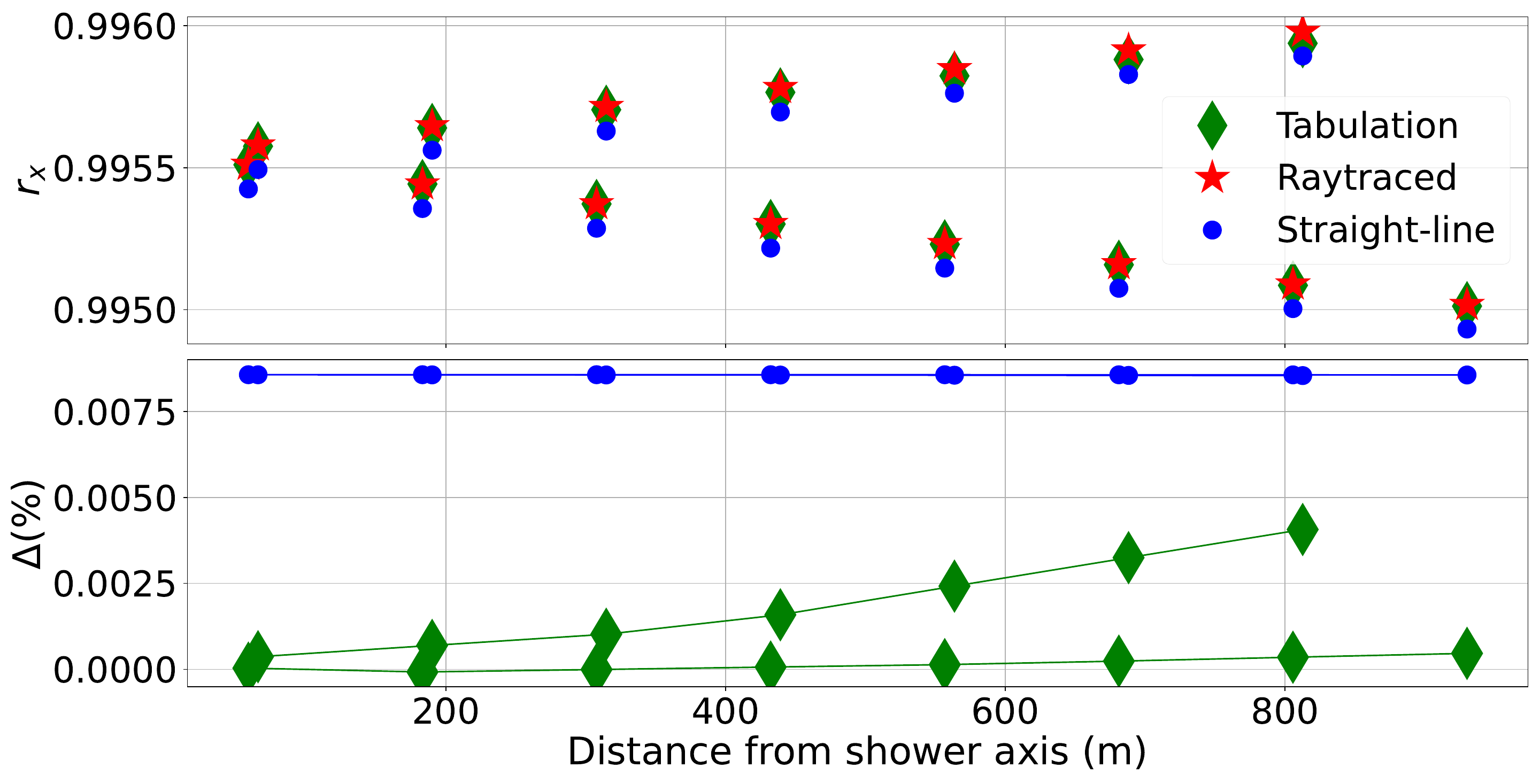}    \caption{A verification for the use of the line model. Top: the horizontal component of unit vectors used in the boostfactor calculation. Red stars are the (correct) values for the launch vector, blue are the values for the straight-line vector ($r_x$) and green is the value reconstructed from our tabulation procedure ($r_{L,x}$). Bottom: the difference between these components as a percentage of the launch vector value: in green the relative difference between the ray traced and the tabulated value, and in blue the relative difference between the straight-line and the ray traced value. Note that the lateral distance from the shower axis was sampled in both directions, causing a doubling of values.}    \label{fig:LineModel}
\end{figure}

A verification of this assumption was made for a geometry of zenith angle $85^{\circ}$, where we traced rays to a region around a typical value of $X_{\mathrm{max}}$, with emitters placed at a lateral distance of around $1$ km from the shower axis. We then compared the horizontal components of:
\begin{itemize}
    \item The straight-line unit vector connecting emitter and receiver,
    \item The initial launch unit vector of the ray connecting emitter and receiver obtained through direct ray tracing, 
    \item The reconstruction of the launch unit vector using tabulation. 
\end{itemize}
The result of this comparison is visualized in Fig.~\ref{fig:LineModel}, where it can be seen that the value from tabulation has good agreement with the correct launch vector value, with a difference of $0.25 \%$  at lateral distances of 500 m, compared to the difference of about $1\%$ when comparing to the straight line value.
This result is interesting because the difference between straight-line  and launch components stays near constant for all lateral distances. Because of this, the tabulation procedure can reproduce the correct value for the horizontal component, as it maps the straight line value to the launch value assuming a position on the shower axis.
\section{Accounting for early-late effects}\label{sec.appendixC}
The accounting for early-late effects in this work follows closely the procedure explained in~\cite{Schluter:2022mhq}. A short summary is given in this appendix.
\\
\par
\begin{figure} 
    \centering
    \includegraphics[width=\linewidth]{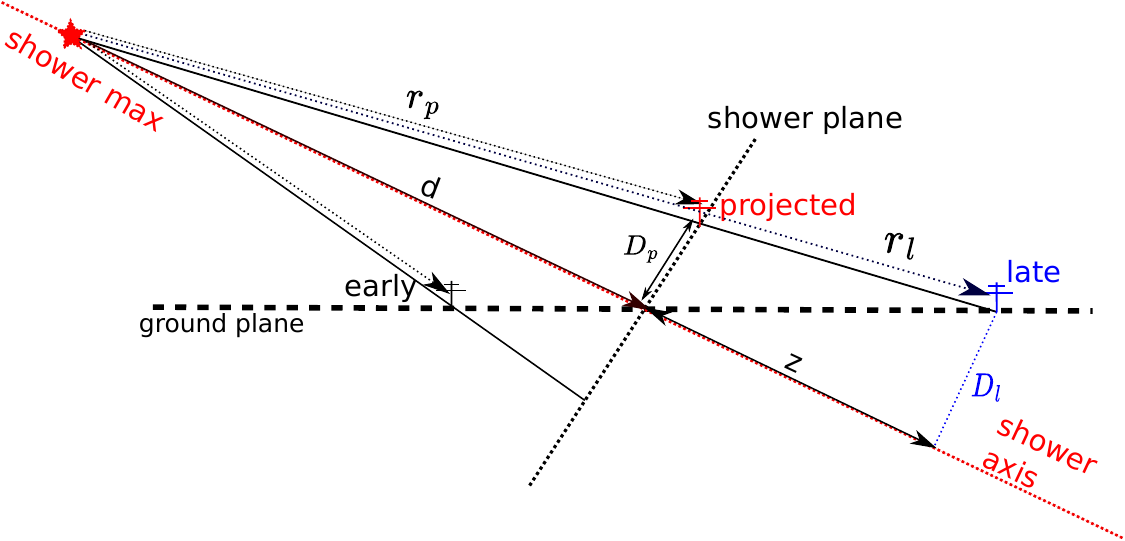}
    \caption{A diagram showing the shower axis of an inclined shower alongside an early and late receiver. Also indicated are the distances used in the derivation of the correction factor used to account for early late effects, as well as the projection on the shower plane of the late receiver position.}
    \label{fig:Early-Late}
\end{figure}

Early late effects arise from a difference in  distance depending on receiver position.
In the context of radiation coming from inclined air shower geometries the associated difference in travel time can be interpreted as a difference in geometrical distance between $X_{\mathrm{max}}$ and the receiver as shown in Fig.~\ref{fig:Early-Late}. A receiver position with a lower associated travel time is often called early, while those with a higher associated travel time are called late receivers.
\\
\par
To quantize and correct early late effects: consider radiation coming from the shower maximum in the form of spherical waves. The amplitude of the electric field then scales as $\frac{1}{r}$. The goal is now to scale the amplitude seen by a receiver to match the amplitude that would be seen by the associated projection in the shower plane.
Consider now the electric field amplitude $E_l$ as seen by the late receiver shown in~\ref{fig:Early-Late}. The relation to the electric field amplitude $E_p$ as seen by the projected receiver in the shower axis is then
\begin{equation}
    E_p=\frac{r_l}{r_p}E_l.
\end{equation}
Since $\frac{r_l}{r_p}=\frac{d+z}{d}$ this can be rewritten as
\begin{align}
    E_\mathrm{p}= F \cdot E_l,
    \\
    F= \frac{d+z}{d}.
\end{align}
Note that z is positive for late receivers and negative for early receivers.
\\
\par
Since fluence scales with the square of the electric field amplitude the correction factor between  the fluence $f_l$ as observed by the late receiver to the associated fluence $f_p$ observed from the projected receiver position is
\begin{align}
    f_p=F^2 \cdot f_l
\end{align}
The derivation of the correction factor for early receivers is identical, the only difference being that $z$ is negative and thus $F<1$. 
\\
\par
The correction factor F can also be used to find the distance of the projected receiver position from the shower axis
\begin{align}
    \frac{D_l}{D_p}=\frac{d+z}{d}=F
\end{align}
It is this $D_p$ which is shown on the horizontal axes of the figures in Fig.~\ref{fig:FluencePlots}.
\section{Geometrical path length and Optical path length} \label{sec.appendixD}
To verify that the electric field amplitude should scale with the inverse of the geometrical distance $R$ and not the optical distance $L$, one can calculate the electric field of a moving charge for a medium with constant index of refraction $n > 1$
\begin{multline}
        \mathbf{E}= -\frac{1}{4\pi\epsilon_0}\frac{qc}{(Rc-n\mathbf{R}\cdot\mathbf{v})^3}\Bigg[ \left(n \mathbf{v} -\frac{\mathbf{v}}{n} + \frac{R}{c} \mathbf{a}\right) \cdot (Rc - n \mathbf{R} \cdot \mathbf{v}) 
        \\
     + \left( \frac{n R}{c}\cdot \mathbf{v} - n^2 \mathbf{R}\right) \left(\frac{c^2}{n^2} -  v^2 +  \mathbf{R}\cdot\mathbf{a} \right)\Bigg],
\end{multline}
where bold symbols represent vectors. $\mathbf{v}$ is the velocity, $\mathbf{a}$ represents the acceleration, $v= |\mathbf{v}|$ and $R=|\mathbf{R}|$. The derivation of this result closely follows the steps detailed in \cite{Griffiths_2017}.
\\
\par
If we now consider the simplest case of a stationary charge $q$, for which both velocity and acceleration terms vanish, we arrive at the following expression for the electric field
\begin{gather*}
    \mathbf{E}=\frac{1}{4\pi \epsilon_0} \frac{q}{c^2} \frac{1}{R^3} \cdot n^2 \mathbf{R} \cdot \frac{c^2}{n^2},
    \\
    \mathbf{E}= \frac{1}{4\pi \epsilon_0} \frac{q}{R^3} \cdot \mathbf{R}.
\end{gather*}
This is the standard Coulomb field, which we find scales with the inverse square of the geometrical distance $\frac{1}{R^2}$. From this result, we conclude that the $R$ in Eq.~\ref{eq:End-point} should represent the geometrical distance $R$ and not the optical path length $L$. 

\bibliography{Bibliography}

\providecommand{\noopsort}[1]{}\providecommand{\singleletter}[1]{#1}%
\begin{thebibliography}{34}%
\makeatletter
\providecommand \@ifxundefined [1]{%
 \@ifx{#1\undefined}
}%
\providecommand \@ifnum [1]{%
 \ifnum #1\expandafter \@firstoftwo
 \else \expandafter \@secondoftwo
 \fi
}%
\providecommand \@ifx [1]{%
 \ifx #1\expandafter \@firstoftwo
 \else \expandafter \@secondoftwo
 \fi
}%
\providecommand \natexlab [1]{#1}%
\providecommand \enquote  [1]{``#1''}%
\providecommand \bibnamefont  [1]{#1}%
\providecommand \bibfnamefont [1]{#1}%
\providecommand \citenamefont [1]{#1}%
\providecommand \href@noop [0]{\@secondoftwo}%
\providecommand \href [0]{\begingroup \@sanitize@url \@href}%
\providecommand \@href[1]{\@@startlink{#1}\@@href}%
\providecommand \@@href[1]{\endgroup#1\@@endlink}%
\providecommand \@sanitize@url [0]{\catcode `\\12\catcode `\$12\catcode `\&12\catcode `\#12\catcode `\^12\catcode `\_12\catcode `\%12\relax}%
\providecommand \@@startlink[1]{}%
\providecommand \@@endlink[0]{}%
\providecommand \url  [0]{\begingroup\@sanitize@url \@url }%
\providecommand \@url [1]{\endgroup\@href {#1}{\urlprefix }}%
\providecommand \urlprefix  [0]{URL }%
\providecommand \Eprint [0]{\href }%
\providecommand \doibase [0]{https://doi.org/}%
\providecommand \selectlanguage [0]{\@gobble}%
\providecommand \bibinfo  [0]{\@secondoftwo}%
\providecommand \bibfield  [0]{\@secondoftwo}%
\providecommand \translation [1]{[#1]}%
\providecommand \BibitemOpen [0]{}%
\providecommand \bibitemStop [0]{}%
\providecommand \bibitemNoStop [0]{.\EOS\space}%
\providecommand \EOS [0]{\spacefactor3000\relax}%
\providecommand \BibitemShut  [1]{\csname bibitem#1\endcsname}%
\let\auto@bib@innerbib\@empty
\bibitem [{\citenamefont {Abreu}\ \emph {et~al.}(2021)\citenamefont {Abreu} \emph {et~al.}}]{PierreAuger:2021hun}%
  \BibitemOpen
  \bibfield  {author} {\bibinfo {author} {\bibfnamefont {P.}~\bibnamefont {Abreu}} \emph {et~al.} (\bibinfo {collaboration} {Pierre Auger}),\ }\bibfield  {title} {\bibinfo {title} {{The energy spectrum of cosmic rays beyond the turn-down around $10^{17}$ eV as measured with the surface detector of the Pierre Auger Observatory}},\ }\href {https://doi.org/10.1140/epjc/s10052-021-09700-w} {\bibfield  {journal} {\bibinfo  {journal} {Eur. Phys. J. C}\ }\textbf {\bibinfo {volume} {81}},\ \bibinfo {pages} {966} (\bibinfo {year} {2021})},\ \Eprint {https://arxiv.org/abs/2109.13400} {arXiv:2109.13400 [astro-ph.HE]} \BibitemShut {NoStop}%
\bibitem [{\citenamefont {Abu-Zayyad}\ \emph {et~al.}(2012)\citenamefont {Abu-Zayyad} \emph {et~al.}}]{TelescopeArray:2012vhh}%
  \BibitemOpen
  \bibfield  {author} {\bibinfo {author} {\bibfnamefont {T.}~\bibnamefont {Abu-Zayyad}} \emph {et~al.} (\bibinfo {collaboration} {Telescope Array}),\ }\bibfield  {title} {\bibinfo {title} {{The Energy Spectrum of Telescope Array's Middle Drum Detector and the Direct Comparison to the High Resolution Fly's Eye Experiment}},\ }\href {https://doi.org/10.1016/j.astropartphys.2012.05.012} {\bibfield  {journal} {\bibinfo  {journal} {Astropart. Phys.}\ }\textbf {\bibinfo {volume} {39-40}},\ \bibinfo {pages} {109} (\bibinfo {year} {2012})},\ \Eprint {https://arxiv.org/abs/1202.5141} {arXiv:1202.5141 [astro-ph.IM]} \BibitemShut {NoStop}%
\bibitem [{\citenamefont {Abbasi}\ \emph {et~al.}(2021)\citenamefont {Abbasi} \emph {et~al.}}]{IceCube:2020wum}%
  \BibitemOpen
  \bibfield  {author} {\bibinfo {author} {\bibfnamefont {R.}~\bibnamefont {Abbasi}} \emph {et~al.} (\bibinfo {collaboration} {IceCube}),\ }\bibfield  {title} {\bibinfo {title} {{The IceCube high-energy starting event sample: Description and flux characterization with 7.5 years of data}},\ }\href {https://doi.org/10.1103/PhysRevD.104.022002} {\bibfield  {journal} {\bibinfo  {journal} {Phys. Rev. D}\ }\textbf {\bibinfo {volume} {104}},\ \bibinfo {pages} {022002} (\bibinfo {year} {2021})},\ \Eprint {https://arxiv.org/abs/2011.03545} {arXiv:2011.03545 [astro-ph.HE]} \BibitemShut {NoStop}%
\bibitem [{\citenamefont {Barrella}\ \emph {et~al.}(2011)\citenamefont {Barrella}, \citenamefont {Barwick},\ and\ \citenamefont {Saltzberg}}]{Barrella:2010vs}%
  \BibitemOpen
  \bibfield  {author} {\bibinfo {author} {\bibfnamefont {T.}~\bibnamefont {Barrella}}, \bibinfo {author} {\bibfnamefont {S.}~\bibnamefont {Barwick}},\ and\ \bibinfo {author} {\bibfnamefont {D.}~\bibnamefont {Saltzberg}},\ }\bibfield  {title} {\bibinfo {title} {{Ross Ice Shelf in situ radio-frequency ice attenuation}},\ }\href {https://doi.org/10.3189/002214311795306691} {\bibfield  {journal} {\bibinfo  {journal} {J. Glaciol.}\ }\textbf {\bibinfo {volume} {57}},\ \bibinfo {pages} {61} (\bibinfo {year} {2011})},\ \Eprint {https://arxiv.org/abs/1011.0477} {arXiv:1011.0477 [astro-ph.IM]} \BibitemShut {NoStop}%
\bibitem [{\citenamefont {Barwick}\ \emph {et~al.}(2005)\citenamefont {Barwick}, \citenamefont {Besson}, \citenamefont {Gorham},\ and\ \citenamefont {Saltzberg}}]{Barwick:2005zz}%
  \BibitemOpen
  \bibfield  {author} {\bibinfo {author} {\bibfnamefont {S.}~\bibnamefont {Barwick}}, \bibinfo {author} {\bibfnamefont {D.}~\bibnamefont {Besson}}, \bibinfo {author} {\bibfnamefont {P.}~\bibnamefont {Gorham}},\ and\ \bibinfo {author} {\bibfnamefont {D.}~\bibnamefont {Saltzberg}},\ }\bibfield  {title} {\bibinfo {title} {{South Polar in situ radio-frequency ice attenuation}},\ }\href {https://doi.org/10.3189/172756505781829467} {\bibfield  {journal} {\bibinfo  {journal} {J. Glaciol.}\ }\textbf {\bibinfo {volume} {51}},\ \bibinfo {pages} {231} (\bibinfo {year} {2005})}\BibitemShut {NoStop}%
\bibitem [{\citenamefont {Besson}\ \emph {et~al.}(2008)\citenamefont {Besson} \emph {et~al.}}]{Besson:2007jja}%
  \BibitemOpen
  \bibfield  {author} {\bibinfo {author} {\bibfnamefont {D.~Z.}\ \bibnamefont {Besson}} \emph {et~al.},\ }\bibfield  {title} {\bibinfo {title} {{In situ radioglaciological measurements near Taylor Dome, Antarctica and implications for UHE neutrino astronomy}},\ }\href {https://doi.org/10.1016/j.astropartphys.2007.12.004} {\bibfield  {journal} {\bibinfo  {journal} {Astropart. Phys.}\ }\textbf {\bibinfo {volume} {29}},\ \bibinfo {pages} {130} (\bibinfo {year} {2008})},\ \Eprint {https://arxiv.org/abs/astro-ph/0703413} {arXiv:astro-ph/0703413} \BibitemShut {NoStop}%
\bibitem [{\citenamefont {Avva}\ \emph {et~al.}(2015)\citenamefont {Avva}, \citenamefont {Kovac}, \citenamefont {Miki}, \citenamefont {Saltzberg},\ and\ \citenamefont {Vieregg}}]{Avva:2014ena}%
  \BibitemOpen
  \bibfield  {author} {\bibinfo {author} {\bibfnamefont {J.}~\bibnamefont {Avva}}, \bibinfo {author} {\bibfnamefont {J.~M.}\ \bibnamefont {Kovac}}, \bibinfo {author} {\bibfnamefont {C.}~\bibnamefont {Miki}}, \bibinfo {author} {\bibfnamefont {D.}~\bibnamefont {Saltzberg}},\ and\ \bibinfo {author} {\bibfnamefont {A.~G.}\ \bibnamefont {Vieregg}},\ }\bibfield  {title} {\bibinfo {title} {{An in situ measurement of the radio-frequency attenuation in ice at Summit Station, Greenland}},\ }\href {https://doi.org/10.3189/2015JoG15J057} {\bibfield  {journal} {\bibinfo  {journal} {J. Glaciol.}\ }\textbf {\bibinfo {volume} {61}},\ \bibinfo {pages} {1005} (\bibinfo {year} {2015})},\ \Eprint {https://arxiv.org/abs/1409.5413} {arXiv:1409.5413 [astro-ph.IM]} \BibitemShut {NoStop}%
\bibitem [{\citenamefont {Huege}(2016)}]{Huege:2016veh}%
  \BibitemOpen
  \bibfield  {author} {\bibinfo {author} {\bibfnamefont {T.}~\bibnamefont {Huege}},\ }\bibfield  {title} {\bibinfo {title} {{Radio detection of cosmic ray air showers in the digital era}},\ }\href {https://doi.org/10.1016/j.physrep.2016.02.001} {\bibfield  {journal} {\bibinfo  {journal} {Phys. Rept.}\ }\textbf {\bibinfo {volume} {620}},\ \bibinfo {pages} {1} (\bibinfo {year} {2016})},\ \Eprint {https://arxiv.org/abs/1601.07426} {arXiv:1601.07426 [astro-ph.IM]} \BibitemShut {NoStop}%
\bibitem [{\citenamefont {Stasielak}(2022)}]{AugerPrimeUpgrade}%
  \BibitemOpen
  \bibfield  {author} {\bibinfo {author} {\bibfnamefont {J.}~\bibnamefont {Stasielak}} (\bibinfo {collaboration} {Pierre Auger}),\ }\bibfield  {title} {\bibinfo {title} {{AugerPrime - The upgrade of the Pierre Auger Observatory}},\ }\href {https://doi.org/10.1142/S0217751X22400127} {\bibfield  {journal} {\bibinfo  {journal} {Int. J. Mod. Phys. A}\ }\textbf {\bibinfo {volume} {37}},\ \bibinfo {pages} {2240012} (\bibinfo {year} {2022})},\ \Eprint {https://arxiv.org/abs/2110.09487} {arXiv:2110.09487 [astro-ph.HE]} \BibitemShut {NoStop}%
\bibitem [{\citenamefont {Huege}(2023)}]{Huege:2023pfb}%
  \BibitemOpen
  \bibfield  {author} {\bibinfo {author} {\bibfnamefont {T.}~\bibnamefont {Huege}} (\bibinfo {collaboration} {Pierre Auger}),\ }\bibfield  {title} {\bibinfo {title} {{The Radio Detector of the Pierre Auger Observatory \textendash{} status and expected performance}},\ }\href {https://doi.org/10.1051/epjconf/202328306002} {\bibfield  {journal} {\bibinfo  {journal} {EPJ Web Conf.}\ }\textbf {\bibinfo {volume} {283}},\ \bibinfo {pages} {06002} (\bibinfo {year} {2023})},\ \Eprint {https://arxiv.org/abs/2305.10104} {arXiv:2305.10104 [astro-ph.IM]} \BibitemShut {NoStop}%
\bibitem [{\citenamefont {\'Alvarez-Mu\~niz}\ \emph {et~al.}(2020)\citenamefont {\'Alvarez-Mu\~niz} \emph {et~al.}}]{GRAND:2018iaj}%
  \BibitemOpen
  \bibfield  {author} {\bibinfo {author} {\bibfnamefont {J.}~\bibnamefont {\'Alvarez-Mu\~niz}} \emph {et~al.} (\bibinfo {collaboration} {GRAND}),\ }\bibfield  {title} {\bibinfo {title} {{The Giant Radio Array for Neutrino Detection (GRAND): Science and Design}},\ }\href {https://doi.org/10.1007/s11433-018-9385-7} {\bibfield  {journal} {\bibinfo  {journal} {Sci. China Phys. Mech. Astron.}\ }\textbf {\bibinfo {volume} {63}},\ \bibinfo {pages} {219501} (\bibinfo {year} {2020})},\ \Eprint {https://arxiv.org/abs/1810.09994} {arXiv:1810.09994 [astro-ph.HE]} \BibitemShut {NoStop}%
\bibitem [{\citenamefont {Southall}\ \emph {et~al.}(2023)\citenamefont {Southall} \emph {et~al.}}]{Southall:2022yil}%
  \BibitemOpen
  \bibfield  {author} {\bibinfo {author} {\bibfnamefont {D.}~\bibnamefont {Southall}} \emph {et~al.},\ }\bibfield  {title} {\bibinfo {title} {{Design and initial performance of the prototype for the BEACON instrument for detection of ultrahigh energy particles}},\ }\href {https://doi.org/10.1016/j.nima.2022.167889} {\bibfield  {journal} {\bibinfo  {journal} {Nucl. Instrum. Meth. A}\ }\textbf {\bibinfo {volume} {1048}},\ \bibinfo {pages} {167889} (\bibinfo {year} {2023})},\ \Eprint {https://arxiv.org/abs/2206.09660} {arXiv:2206.09660 [astro-ph.IM]} \BibitemShut {NoStop}%
\bibitem [{\citenamefont {Nam}(2016)}]{TAROGE}%
  \BibitemOpen
  \bibfield  {author} {\bibinfo {author} {\bibfnamefont {J.}~\bibnamefont {Nam}},\ }\bibfield  {title} {\bibinfo {title} {{Taiwan Astroparticle Radiowave Observatory for Geo-synchrotron Emissions (TAROGE)}},\ }\href {https://doi.org/10.22323/1.236.0663} {\bibfield  {journal} {\bibinfo  {journal} {PoS}\ }\textbf {\bibinfo {volume} {ICRC2015}},\ \bibinfo {pages} {663} (\bibinfo {year} {2016})}\BibitemShut {NoStop}%
\bibitem [{\citenamefont {Wang}\ \emph {et~al.}(2022)\citenamefont {Wang} \emph {et~al.}}]{TAROGE-M:2022soh}%
  \BibitemOpen
  \bibfield  {author} {\bibinfo {author} {\bibfnamefont {S.-H.}\ \bibnamefont {Wang}} \emph {et~al.} (\bibinfo {collaboration} {TAROGE, Arianna}),\ }\bibfield  {title} {\bibinfo {title} {{TAROGE-M: radio antenna array on antarctic high mountain for detecting near-horizontal ultra-high energy air showers}},\ }\href {https://doi.org/10.1088/1475-7516/2022/11/022} {\bibfield  {journal} {\bibinfo  {journal} {JCAP}\ }\textbf {\bibinfo {volume} {11}},\ \bibinfo {pages} {022}},\ \Eprint {https://arxiv.org/abs/2207.10616} {arXiv:2207.10616 [astro-ph.HE]} \BibitemShut {NoStop}%
\bibitem [{\citenamefont {Huege}\ \emph {et~al.}(2013)\citenamefont {Huege}, \citenamefont {Ludwig},\ and\ \citenamefont {James}}]{Huege:2013vt}%
  \BibitemOpen
  \bibfield  {author} {\bibinfo {author} {\bibfnamefont {T.}~\bibnamefont {Huege}}, \bibinfo {author} {\bibfnamefont {M.}~\bibnamefont {Ludwig}},\ and\ \bibinfo {author} {\bibfnamefont {C.~W.}\ \bibnamefont {James}},\ }\bibfield  {title} {\bibinfo {title} {{Simulating radio emission from air showers with CoREAS}},\ }\href {https://doi.org/10.1063/1.4807534} {\bibfield  {journal} {\bibinfo  {journal} {AIP Conf. Proc.}\ }\textbf {\bibinfo {volume} {1535}},\ \bibinfo {pages} {128} (\bibinfo {year} {2013})},\ \Eprint {https://arxiv.org/abs/1301.2132} {arXiv:1301.2132 [astro-ph.HE]} \BibitemShut {NoStop}%
\bibitem [{\citenamefont {Alvarez-Muniz}\ \emph {et~al.}(2012)\citenamefont {Alvarez-Muniz}, \citenamefont {Carvalho}, \citenamefont {Tueros},\ and\ \citenamefont {Zas}}]{Alvarez-Muniz:2010hbb}%
  \BibitemOpen
  \bibfield  {author} {\bibinfo {author} {\bibfnamefont {J.}~\bibnamefont {Alvarez-Muniz}}, \bibinfo {author} {\bibfnamefont {W.~R.}\ \bibnamefont {Carvalho}, \bibfnamefont {Jr.}}, \bibinfo {author} {\bibfnamefont {M.}~\bibnamefont {Tueros}},\ and\ \bibinfo {author} {\bibfnamefont {E.}~\bibnamefont {Zas}},\ }\bibfield  {title} {\bibinfo {title} {{Coherent Cherenkov radio pulses from hadronic showers up to EeV energies}},\ }\href {https://doi.org/10.1016/j.astropartphys.2011.10.002} {\bibfield  {journal} {\bibinfo  {journal} {Astropart. Phys.}\ }\textbf {\bibinfo {volume} {35}},\ \bibinfo {pages} {287} (\bibinfo {year} {2012})},\ \Eprint {https://arxiv.org/abs/1005.0552} {arXiv:1005.0552 [astro-ph.HE]} \BibitemShut {NoStop}%
\bibitem [{\citenamefont {Alvarez-Mu\~niz}\ \emph {et~al.}(2015)\citenamefont {Alvarez-Mu\~niz}, \citenamefont {Carvalho}, \citenamefont {Garc\'\i{}a-Fern\'andez}, \citenamefont {Schoorlemmer},\ and\ \citenamefont {Zas}}]{Alvarez-Muniz:2015ayz}%
  \BibitemOpen
  \bibfield  {author} {\bibinfo {author} {\bibfnamefont {J.}~\bibnamefont {Alvarez-Mu\~niz}}, \bibinfo {author} {\bibfnamefont {W.~R.}\ \bibnamefont {Carvalho}}, \bibinfo {author} {\bibfnamefont {D.}~\bibnamefont {Garc\'\i{}a-Fern\'andez}}, \bibinfo {author} {\bibfnamefont {H.}~\bibnamefont {Schoorlemmer}},\ and\ \bibinfo {author} {\bibfnamefont {E.}~\bibnamefont {Zas}},\ }\bibfield  {title} {\bibinfo {title} {{Simulations of reflected radio signals from cosmic ray induced air showers}},\ }\href {https://doi.org/10.1016/j.astropartphys.2014.12.005} {\bibfield  {journal} {\bibinfo  {journal} {Astropart. Phys.}\ }\textbf {\bibinfo {volume} {66}},\ \bibinfo {pages} {31} (\bibinfo {year} {2015})},\ \Eprint {https://arxiv.org/abs/1502.02117} {arXiv:1502.02117 [astro-ph.HE]} \BibitemShut {NoStop}%
\bibitem [{\citenamefont {Schl\"uter}\ \emph {et~al.}(2020)\citenamefont {Schl\"uter}, \citenamefont {Gottowik}, \citenamefont {Huege},\ and\ \citenamefont {Rautenberg}}]{Schluter:2020tdz}%
  \BibitemOpen
  \bibfield  {author} {\bibinfo {author} {\bibfnamefont {F.}~\bibnamefont {Schl\"uter}}, \bibinfo {author} {\bibfnamefont {M.}~\bibnamefont {Gottowik}}, \bibinfo {author} {\bibfnamefont {T.}~\bibnamefont {Huege}},\ and\ \bibinfo {author} {\bibfnamefont {J.}~\bibnamefont {Rautenberg}},\ }\bibfield  {title} {\bibinfo {title} {{Refractive displacement of the radio-emission footprint of inclined air showers simulated with CoREAS}},\ }\href {https://doi.org/10.1140/epjc/s10052-020-8216-z} {\bibfield  {journal} {\bibinfo  {journal} {Eur. Phys. J. C}\ }\textbf {\bibinfo {volume} {80}},\ \bibinfo {pages} {643} (\bibinfo {year} {2020})},\ \Eprint {https://arxiv.org/abs/2005.06775} {arXiv:2005.06775 [astro-ph.IM]} \BibitemShut {NoStop}%
\bibitem [{\citenamefont {Werner}\ and\ \citenamefont {Scholten}(2008)}]{Werner:2007kh}%
  \BibitemOpen
  \bibfield  {author} {\bibinfo {author} {\bibfnamefont {K.}~\bibnamefont {Werner}}\ and\ \bibinfo {author} {\bibfnamefont {O.}~\bibnamefont {Scholten}},\ }\bibfield  {title} {\bibinfo {title} {{Macroscopic Treatment of Radio Emission from Cosmic Ray Air Showers based on Shower Simulations}},\ }\href {https://doi.org/10.1016/j.astropartphys.2008.04.004} {\bibfield  {journal} {\bibinfo  {journal} {Astropart. Phys.}\ }\textbf {\bibinfo {volume} {29}},\ \bibinfo {pages} {393} (\bibinfo {year} {2008})},\ \Eprint {https://arxiv.org/abs/0712.2517} {arXiv:0712.2517 [astro-ph]} \BibitemShut {NoStop}%
\bibitem [{\citenamefont {Deaconu}\ \emph {et~al.}(2018)\citenamefont {Deaconu}, \citenamefont {Vieregg}, \citenamefont {Wissel}, \citenamefont {Bowen}, \citenamefont {Chipman}, \citenamefont {Gupta}, \citenamefont {Miki}, \citenamefont {Nichol},\ and\ \citenamefont {Saltzberg}}]{Deaconu:2018bkf}%
  \BibitemOpen
  \bibfield  {author} {\bibinfo {author} {\bibfnamefont {C.}~\bibnamefont {Deaconu}}, \bibinfo {author} {\bibfnamefont {A.~G.}\ \bibnamefont {Vieregg}}, \bibinfo {author} {\bibfnamefont {S.~A.}\ \bibnamefont {Wissel}}, \bibinfo {author} {\bibfnamefont {J.}~\bibnamefont {Bowen}}, \bibinfo {author} {\bibfnamefont {S.}~\bibnamefont {Chipman}}, \bibinfo {author} {\bibfnamefont {A.}~\bibnamefont {Gupta}}, \bibinfo {author} {\bibfnamefont {C.}~\bibnamefont {Miki}}, \bibinfo {author} {\bibfnamefont {R.~J.}\ \bibnamefont {Nichol}},\ and\ \bibinfo {author} {\bibfnamefont {D.}~\bibnamefont {Saltzberg}},\ }\bibfield  {title} {\bibinfo {title} {{Measurements and Modeling of Near-Surface Radio Propagation in Glacial Ice and Implications for Neutrino Experiments}},\ }\href {https://doi.org/10.1103/PhysRevD.98.043010} {\bibfield  {journal} {\bibinfo  {journal} {Phys. Rev. D}\ }\textbf {\bibinfo {volume} {98}},\ \bibinfo {pages} {043010} (\bibinfo {year} {2018})},\ \Eprint {https://arxiv.org/abs/1805.12576} {arXiv:1805.12576
  [astro-ph.IM]} \BibitemShut {NoStop}%
\bibitem [{\citenamefont {Prohira}\ \emph {et~al.}(2021)\citenamefont {Prohira} \emph {et~al.}}]{RadarEchoTelescope:2020nhe}%
  \BibitemOpen
  \bibfield  {author} {\bibinfo {author} {\bibfnamefont {S.}~\bibnamefont {Prohira}} \emph {et~al.} (\bibinfo {collaboration} {Radar Echo Telescope}),\ }\bibfield  {title} {\bibinfo {title} {{Modeling in-ice radio propagation with parabolic equation methods}},\ }\href {https://doi.org/10.1103/PhysRevD.103.103007} {\bibfield  {journal} {\bibinfo  {journal} {Phys. Rev. D}\ }\textbf {\bibinfo {volume} {103}},\ \bibinfo {pages} {103007} (\bibinfo {year} {2021})},\ \Eprint {https://arxiv.org/abs/2011.05997} {arXiv:2011.05997 [astro-ph.IM]} \BibitemShut {NoStop}%
\bibitem [{\citenamefont {Windischhofer}\ \emph {et~al.}(2023)\citenamefont {Windischhofer}, \citenamefont {Welling},\ and\ \citenamefont {Deaconu}}]{Windischhofer:2023ahw}%
  \BibitemOpen
  \bibfield  {author} {\bibinfo {author} {\bibfnamefont {P.}~\bibnamefont {Windischhofer}}, \bibinfo {author} {\bibfnamefont {C.}~\bibnamefont {Welling}},\ and\ \bibinfo {author} {\bibfnamefont {C.}~\bibnamefont {Deaconu}},\ }\bibfield  {title} {\bibinfo {title} {{Eisvogel: Exact and efficient calculations of radio emissions from in-ice neutrino showers}},\ }\href {https://doi.org/10.22323/1.444.1157} {\bibfield  {journal} {\bibinfo  {journal} {PoS}\ }\textbf {\bibinfo {volume} {ICRC2023}},\ \bibinfo {pages} {1157} (\bibinfo {year} {2023})}\BibitemShut {NoStop}%
\bibitem [{\citenamefont {\textit{et al}}(2019)}]{RNO-G}%
  \BibitemOpen
  \bibfield  {author} {\bibinfo {author} {\bibfnamefont {J.~A.~A.}\ \bibnamefont {\textit{et al}}},\ }\href@noop {} {\bibinfo {title} {The next-generation radio neutrino observatory -- multi-messenger neutrino astrophysics at extreme energies}} (\bibinfo {year} {2019}),\ \Eprint {https://arxiv.org/abs/1907.12526} {arXiv:1907.12526 [astro-ph.HE]} \BibitemShut {NoStop}%
\bibitem [{\citenamefont {De~Kockere}\ \emph {et~al.}(2024)\citenamefont {De~Kockere}, \citenamefont {Van~den Broeck}, \citenamefont {Latif}, \citenamefont {de~Vries}, \citenamefont {van Eijndhoven}, \citenamefont {Huege},\ and\ \citenamefont {Buitink}}]{DeKockere:2024qmc}%
  \BibitemOpen
  \bibfield  {author} {\bibinfo {author} {\bibfnamefont {S.}~\bibnamefont {De~Kockere}}, \bibinfo {author} {\bibfnamefont {D.}~\bibnamefont {Van~den Broeck}}, \bibinfo {author} {\bibfnamefont {U.~A.}\ \bibnamefont {Latif}}, \bibinfo {author} {\bibfnamefont {K.~D.}\ \bibnamefont {de~Vries}}, \bibinfo {author} {\bibfnamefont {N.}~\bibnamefont {van Eijndhoven}}, \bibinfo {author} {\bibfnamefont {T.}~\bibnamefont {Huege}},\ and\ \bibinfo {author} {\bibfnamefont {S.}~\bibnamefont {Buitink}},\ }\bibfield  {title} {\bibinfo {title} {{Simulation of radio signals from cosmic-ray cascades in air and ice as observed by in-ice Askaryan radio detectors}},\ }\href {https://doi.org/10.1103/PhysRevD.110.023010} {\bibfield  {journal} {\bibinfo  {journal} {Phys. Rev. D}\ }\textbf {\bibinfo {volume} {110}},\ \bibinfo {pages} {023010} (\bibinfo {year} {2024})},\ \Eprint {https://arxiv.org/abs/2403.15358} {arXiv:2403.15358 [astro-ph.HE]} \BibitemShut {NoStop}%
\bibitem [{\citenamefont {James}\ \emph {et~al.}(2011)\citenamefont {James}, \citenamefont {Falcke}, \citenamefont {Huege},\ and\ \citenamefont {Ludwig}}]{EndPoints}%
  \BibitemOpen
  \bibfield  {author} {\bibinfo {author} {\bibfnamefont {C.~W.}\ \bibnamefont {James}}, \bibinfo {author} {\bibfnamefont {H.}~\bibnamefont {Falcke}}, \bibinfo {author} {\bibfnamefont {T.}~\bibnamefont {Huege}},\ and\ \bibinfo {author} {\bibfnamefont {M.}~\bibnamefont {Ludwig}},\ }\bibfield  {title} {\bibinfo {title} {General description of electromagnetic radiation processes based on instantaneous charge acceleration in ``endpoints''},\ }\href {https://doi.org/10.1103/PhysRevE.84.056602} {\bibfield  {journal} {\bibinfo  {journal} {Phys. Rev. E}\ }\textbf {\bibinfo {volume} {84}},\ \bibinfo {pages} {056602} (\bibinfo {year} {2011})}\BibitemShut {NoStop}%
\bibitem [{\citenamefont {Ludwig}\ and\ \citenamefont {Huege}(2011)}]{Ludwig:2010pf}%
  \BibitemOpen
  \bibfield  {author} {\bibinfo {author} {\bibfnamefont {M.}~\bibnamefont {Ludwig}}\ and\ \bibinfo {author} {\bibfnamefont {T.}~\bibnamefont {Huege}},\ }\bibfield  {title} {\bibinfo {title} {{REAS3: Monte Carlo simulations of radio emission from cosmic ray air showers using an 'end-point' formalism}},\ }\href {https://doi.org/10.1016/j.astropartphys.2010.10.012} {\bibfield  {journal} {\bibinfo  {journal} {Astropart. Phys.}\ }\textbf {\bibinfo {volume} {34}},\ \bibinfo {pages} {438} (\bibinfo {year} {2011})},\ \Eprint {https://arxiv.org/abs/1010.5343} {arXiv:1010.5343 [astro-ph.HE]} \BibitemShut {NoStop}%
\bibitem [{\citenamefont {Holm}(2011)}]{holm2011geometric}%
  \BibitemOpen
  \bibfield  {author} {\bibinfo {author} {\bibfnamefont {D.}~\bibnamefont {Holm}},\ }\href {https://books.google.be/books?id=RM42DwAAQBAJ} {\emph {\bibinfo {title} {Geometric Mechanics - Part I: Dynamics And Symmetry (2nd Edition)}}}\ (\bibinfo  {publisher} {World Scientific Publishing Company},\ \bibinfo {year} {2011})\BibitemShut {NoStop}%
\bibitem [{\citenamefont {D.~Heck}\ and\ \citenamefont {Pierog}(2024)}]{CorsikaWebsite}%
  \BibitemOpen
  \bibfield  {author} {\bibinfo {author} {\bibfnamefont {T.~H.}\ \bibnamefont {D.~Heck}}\ and\ \bibinfo {author} {\bibfnamefont {T.}~\bibnamefont {Pierog}},\ }\href@noop {} {\bibinfo {title} {Corsika documentation}} (\bibinfo {year} {2024}),\ \bibinfo {note} {available at: \url{https://www.iap.kit.edu/corsika/70.php}}\BibitemShut {NoStop}%
\bibitem [{\citenamefont {Lafebre}\ \emph {et~al.}(2009)\citenamefont {Lafebre}, \citenamefont {Engel}, \citenamefont {Falcke}, \citenamefont {Horandel}, \citenamefont {Huege}, \citenamefont {Kuijpers},\ and\ \citenamefont {Ulrich}}]{Lafebre:2009en}%
  \BibitemOpen
  \bibfield  {author} {\bibinfo {author} {\bibfnamefont {S.}~\bibnamefont {Lafebre}}, \bibinfo {author} {\bibfnamefont {R.}~\bibnamefont {Engel}}, \bibinfo {author} {\bibfnamefont {H.}~\bibnamefont {Falcke}}, \bibinfo {author} {\bibfnamefont {J.}~\bibnamefont {Horandel}}, \bibinfo {author} {\bibfnamefont {T.}~\bibnamefont {Huege}}, \bibinfo {author} {\bibfnamefont {J.}~\bibnamefont {Kuijpers}},\ and\ \bibinfo {author} {\bibfnamefont {R.}~\bibnamefont {Ulrich}},\ }\bibfield  {title} {\bibinfo {title} {{Universality of electron-positron distributions in extensive air showers}},\ }\href {https://doi.org/10.1016/j.astropartphys.2009.02.002} {\bibfield  {journal} {\bibinfo  {journal} {Astropart. Phys.}\ }\textbf {\bibinfo {volume} {31}},\ \bibinfo {pages} {243} (\bibinfo {year} {2009})},\ \Eprint {https://arxiv.org/abs/0902.0548} {arXiv:0902.0548 [astro-ph.HE]} \BibitemShut {NoStop}%
\bibitem [{\citenamefont {Aab}\ \emph {et~al.}(2016)\citenamefont {Aab} \emph {et~al.}}]{PierreAuger:2015hbf}%
  \BibitemOpen
  \bibfield  {author} {\bibinfo {author} {\bibfnamefont {A.}~\bibnamefont {Aab}} \emph {et~al.} (\bibinfo {collaboration} {Pierre Auger}),\ }\bibfield  {title} {\bibinfo {title} {{Energy Estimation of Cosmic Rays with the Engineering Radio Array of the Pierre Auger Observatory}},\ }\href {https://doi.org/10.1103/PhysRevD.93.122005} {\bibfield  {journal} {\bibinfo  {journal} {Phys. Rev. D}\ }\textbf {\bibinfo {volume} {93}},\ \bibinfo {pages} {122005} (\bibinfo {year} {2016})},\ \Eprint {https://arxiv.org/abs/1508.04267} {arXiv:1508.04267 [astro-ph.HE]} \BibitemShut {NoStop}%
\bibitem [{\citenamefont {Schl\"uter}\ and\ \citenamefont {Huege}(2023)}]{Schluter:2022mhq}%
  \BibitemOpen
  \bibfield  {author} {\bibinfo {author} {\bibfnamefont {F.}~\bibnamefont {Schl\"uter}}\ and\ \bibinfo {author} {\bibfnamefont {T.}~\bibnamefont {Huege}},\ }\bibfield  {title} {\bibinfo {title} {{Signal model and event reconstruction for the radio detection of inclined air showers}},\ }\href {https://doi.org/10.1088/1475-7516/2023/01/008} {\bibfield  {journal} {\bibinfo  {journal} {JCAP}\ }\textbf {\bibinfo {volume} {01}},\ \bibinfo {pages} {008}},\ \Eprint {https://arxiv.org/abs/2203.04364} {arXiv:2203.04364 [astro-ph.HE]} \BibitemShut {NoStop}%
\bibitem [{\citenamefont {Scholten}\ \emph {et~al.}(2019)\citenamefont {Scholten}, \citenamefont {Trinh}, \citenamefont {de~Vries},\ and\ \citenamefont {Hare}}]{Scholten:2019kxcMGMR}%
  \BibitemOpen
  \bibfield  {author} {\bibinfo {author} {\bibfnamefont {O.}~\bibnamefont {Scholten}}, \bibinfo {author} {\bibfnamefont {G.}~\bibnamefont {Trinh}}, \bibinfo {author} {\bibfnamefont {K.~D.}\ \bibnamefont {de~Vries}},\ and\ \bibinfo {author} {\bibfnamefont {B.}~\bibnamefont {Hare}},\ }\bibfield  {title} {\bibinfo {title} {{MGMR3D, a semi-analytic code for the obtaining the radio footprint from the shower currents}},\ }\href {https://doi.org/10.1051/epjconf/201921603003} {\bibfield  {journal} {\bibinfo  {journal} {EPJ Web Conf.}\ }\textbf {\bibinfo {volume} {216}},\ \bibinfo {pages} {03003} (\bibinfo {year} {2019})}\BibitemShut {NoStop}%
\bibitem [{\citenamefont {de~Vries}\ \emph {et~al.}(2013)\citenamefont {de~Vries}, \citenamefont {Scholten},\ and\ \citenamefont {Werner}}]{deVries:2013bonEVA}%
  \BibitemOpen
  \bibfield  {author} {\bibinfo {author} {\bibfnamefont {K.~D.}\ \bibnamefont {de~Vries}}, \bibinfo {author} {\bibfnamefont {O.}~\bibnamefont {Scholten}},\ and\ \bibinfo {author} {\bibfnamefont {K.}~\bibnamefont {Werner}},\ }\bibfield  {title} {\bibinfo {title} {{The EVA code; macroscopic modeling of radio emission from air showers based on full MC simulations including a realistic index of refraction}},\ }\href {https://doi.org/10.1063/1.4807535} {\bibfield  {journal} {\bibinfo  {journal} {AIP Conf. Proc.}\ }\textbf {\bibinfo {volume} {1535}},\ \bibinfo {pages} {133} (\bibinfo {year} {2013})}\BibitemShut {NoStop}%
\bibitem [{\citenamefont {Griffiths}(2017)}]{Griffiths_2017}%
  \BibitemOpen
  \bibfield  {author} {\bibinfo {author} {\bibfnamefont {D.~J.}\ \bibnamefont {Griffiths}},\ }\href@noop {} {\emph {\bibinfo {title} {Introduction to Electrodynamics}}},\ \bibinfo {edition} {4th}\ ed.\ (\bibinfo  {publisher} {Cambridge University Press},\ \bibinfo {year} {2017})\BibitemShut {NoStop}%
\end{thebibliography}%
\end{document}